\documentclass[nobibnotes, twocolumn, aps, prb, superscriptaddress]{revtex4-2}
\usepackage[utf8]{inputenc}
\usepackage[letterpaper, margin=1.8cm]{geometry}
\usepackage{graphicx}% Include figure files
\usepackage{dcolumn}% Align table columns on decimal point
\usepackage{bm}% bold math
\usepackage{amsmath}
\usepackage[english]{babel}	
\usepackage[T1]{fontenc}
\usepackage{verbatim}
\usepackage{physics}
\usepackage{mathtools}
\usepackage{booktabs}
\usepackage{gensymb}
\usepackage{amssymb}
\usepackage{color}
\usepackage{bbold} %doubled ones for identity matrix

\usepackage{textcomp}

\bibpunct{[}{]}{,}{s}{,}{,} % 4th arg: "n" for numeric, "s" for numeric superscript, "any letter" for author-year

\usepackage{hyperref} % PDF hyperlinks
\definecolor{dullmagenta}{rgb}{0.4,0,0.4} % #660066
\definecolor{darkblue}{rgb}{0,0,0.4}
\definecolor{medblue}{rgb}{0,0,0.6}
\definecolor{lightblue}{rgb}{0,0,0.8}
\hypersetup{
	colorlinks=true, 		% colorise les liens [true/false] for print
	breaklinks=true,		% permet le retour la ligne dans les liens trop longs 
	linkcolor=lightblue,
	citecolor=lightblue,
	urlcolor=lightblue,
	filecolor=dullmagenta
}
\usepackage[all]{hypcap} % For going to the top of an image when a figure reference is clicked. Use with RevTEX because cannot use the caption package witch does that also.
\graphicspath{{Figs/}{SuppFigs/}}

\makeatletter
\newcommand\setcurrentname[1]{\def\@currentlabelname{#1}}
\def\maketitle{
\@author@finish
\title@column\titleblock@produce
\suppressfloats[t]}
\makeatother

\DeclareUnicodeCharacter{03C0}{$\pi$}
\DeclareUnicodeCharacter{03BC}{$\mu$}

\newcommand{\labelEq}[1]{\label{eq:#1}}
\newcommand{\labelFig}[1]{\label{fig:#1}}
\newcommand{\labelSuppFig}[1]{\label{sfig:#1}}
\newcommand{\labelTab}[1]{\label{tab:#1}}
\newcommand{\labelSuppTab}[1]{\label{stab:#1}}
\newcommand{\labelSec}[1]{\label{sec:#1}}

\newcommand{\refEq}[1]{Eq.~\ref{eq:#1}}
\newcommand{\refFig}[1]{Fig.~\ref{fig:#1}}
\newcommand{\refSuppFig}[1]{Supp.~Fig.~\ref{sfig:#1}}
\newcommand{\refTab}[1]{Tab.~\ref{tab:#1}}
\newcommand{\refSuppTab}[1]{Supp.~Tab.~\ref{stab:#1}}
\newcommand{\refSec}[1]{Section~\ref{sec:#1}}

\usepackage{xcolor}
\usepackage{siunitx}
\usepackage{multirow}
\usepackage{colortbl}
\usepackage{booktabs}

\definecolor{darkgrey}{HTML}{333333}
\definecolor{Q1brown}{HTML}{8C584E}
\definecolor{Q2pink}{HTML}{D676AB}
\definecolor{lightgrey}{HTML}{EFEFEF}

\newcommand{\Vb}{V_{\rm B}}
\newcommand{\Ve}{V_{\varepsilon}}
\newcommand{\B}{\vec{B}}

\newcommand{\g}{\tensor{g}}
\newcommand{\R}{\tensor{R}}

\newcommand{\dg}{\Delta g}
\newcommand{\Edg}{E_{\Delta g}}
\newcommand{\qi}{_{\rm Q1}}
\newcommand{\qii}{_{\rm Q2}}
\newcommand{\qiu}{_{\rm{Q1},\uparrow}}
\newcommand{\qiiu}{_{\rm{Q2},\uparrow}}
\newcommand{\qid}{_{\rm{Q1},\downarrow}}
\newcommand{\qiid}{_{\rm{Q2},\downarrow}}
\newcommand{\SO}{_{\rm{SO}}}
\newcommand{\Jpara}{J_{\parallel}}
\newcommand{\Jperp}{J_{\perp}}
\newcommand{\kd}{\ket{\downarrow}}
\newcommand{\ku}{\ket{\uparrow}}
\newcommand{\kdd}{\ket{\downarrow\downarrow}}
\newcommand{\kdu}{\ket{\downarrow\uparrow}}
\newcommand{\kud}{\ket{\uparrow\downarrow}}
\newcommand{\kuu}{\ket{\uparrow\uparrow}}

\newcommand{\nso}{\hat{n}_{\rm{SO}}}
\newcommand{\muB}{\mu_{\rm{B}}}

\newcommand{\Pauliveci}{\vec{\sigma}_1}
\newcommand{\Paulivecii}{\vec{\sigma}_2}

\newcommand{\gi}{\tensor{g}_1}
\newcommand{\gii}{\tensor{g}_2}

\newcommand{\nvi}{\hat{n}_1}
\newcommand{\nvii}{\hat{n}_2}

\begin{document}
%TC:ignore

\title{Engineering two-qubit gates via anisotropic exchange in germanium spin qubits}

\author{L.~Massai}
\email{l.massai@tudelft.nl}
\author{B.~Het\'enyi}
\author{E.~G.~Kelly}
\author{I.~Seidler}
\author{K.~Tsoukalas}
\author{M.~Aldeghi}
\author{A.~Orekhov}
\author{L.~Sommer}
\author{M.~Pita-Vidal}
\author{U.~von~L\"upke}
\author{S.~Paredes}
\affiliation{IBM Research Europe~-~Zurich, Säumerstrasse 4, 8803 Rüschlikon, Switzerland}
\author{S.W.~Bedell}
\affiliation{IBM Quantum, T.J. Watson Research Center, 1101 Kitchawan Road, Yorktown Heights, New York 10598, USA}
\author{F.J.~Schupp}
\author{M.~Mergenthaler}
\author{G.~Salis}
\author{A.~Fuhrer}
\author{P.~Harvey-Collard}
\email{phc@zurich.ibm.com}
\affiliation{IBM Research Europe~-~Zurich, Säumerstrasse 4, 8803 Rüschlikon, Switzerland}

\date{\today}

\begin{abstract}
Germanium hole spin qubits are a promising and versatile platform for quantum computation and simulation. In this system, strong spin-orbit interaction (SOI) renders the single-qubit $g$-tensor anisotropic and electrically tunable, enabling operational sweet spots with reduced noise sensitivity. SOI also transforms the isotropic two-qubit exchange coupling into an anisotropic tensor whose geometry is inherited from the single-qubit $g$-tensors and spin-flip tunnelling. Here, using two hole spin qubits in a strained-germanium quantum well and full vector control of the magnetic field, we map this exchange tensor, separate it into longitudinal and transverse components, and show that they govern controlled-phase and SWAP-like dynamics, respectively. We find that the longitudinal exchange can be tuned via the magnetic field orientation from a conventional positive value, through zero, to an effectively negative one, as measured by inverted exchange-split spin transitions. The magnetic field direction thus provides continuous control over the interaction Hamiltonian: at a point of purely transverse exchange, we engineer a single-pulse baseband iSWAP, unattainable under isotropic exchange. Linking $g$-tensor geometry to exchange anisotropy establishes native Hamiltonian engineering, enabling spin-based quantum simulation and gate sets selected by the global field orientation alone.

\end{abstract}

\maketitle
%TC:endignore

\section{\labelSec{intro}Introduction}
Hole spins in semiconductor quantum dots (QDs) have recently matured as a versatile platform for quantum information processing\cite{jirovec_singlettriplet_2021, hendrickx_fourqubit_2021, borsoi_shared_2024, john_robust_2025, dijkema_simultaneous_2026} and quantum simulations\cite{hsiao_exciton_2024, jirovec_manybody_2026}. The strong intrinsic spin-orbit interaction (SOI) in these systems gives rise to  electrically-tunable and magnetic-field-dependent features, such as spin-flip tunnelling and anisotropic $g$-tensors. In both silicon and germanium, these properties have been leveraged to demonstrate all-electrical fast control\cite{maurand_cmos_2016, carballido_compromisefree_2025, wang_operating_2024}, noise-resilient operational sweet spots\cite{hendrickx_sweetspot_2024, piot_single_2022, bassi_optimal_2025}, and high-fidelity initialisation and readout\cite{kelly_identifying_2025}.

While the consequences of SOI on single-qubit properties are increasingly well understood\cite{crippa_electrical_2018, abadillo-uriel_holespin_2023, brickson_using_2024, valvo_electrically_2025,  sommer_disentangling_2026}, its influence on the two-qubit exchange interaction is still largely unexplored. The canonical isotropic Heisenberg exchange, $H = \frac{1}{4}J_0~\vec{\sigma}_1 \cdot \vec{\sigma}_2$, where $J_0$ is a real constant, remains valid when the spin operators $\vec{\sigma}_1,\vec{\sigma}_2$ are defined in the same frame. However, strong SOI can complicate this picture: spin-flip tunnelling and site-dependent $g$-tensors make the qubits' Larmor vectors non-parallel and dependent on the magnetic field $\B$. Hence, in their respective eigenbases, the exchange coupling becomes an anisotropic tensor, $\tensor{J}(\B)$, giving rise to distinct longitudinal ($\Jpara$) and transverse ($\Jperp$) components\cite{geyer_anisotropic_2024}. Anisotropies in the exchange interaction can be beneficial when leveraged or detrimental when overlooked, as the ratio of the longitudinal and transverse components dictates the best native two-qubit gate to be implemented in the system, influencing the choice between controlled-phase-like gates~\cite{veldhorst_twoqubit_2015, xue_quantum_2022} and swap-like gates~\cite{petta_coherent_2005, mills_twoqubit_2022, ni_swap_2025}. Understanding and controlling this ratio is therefore essential for optimising gate fidelity and minimising unwanted crosstalk. Anisotropic exchange has been recently measured for holes in silicon finFETs\cite{fuhrer_spin_2022,geyer_anisotropic_2024}, revealing to be mostly dominated by spin-flip tunnelling. In germanium, anisotropic exchange has been indirectly observed in singlet-triplet qubits\cite{saez-mollejo_exchange_2025}, however its effects on Loss-DiVincenzo single-hole spin qubits and two-qubit gates have not been directly measured yet.

In this work, we experimentally investigate the anisotropic exchange interaction between two hole spin qubits in a Ge/SiGe heterostructure. By performing state spectroscopy under full vector control of the magnetic field $\B$, we map the longitudinal $\Jpara$ and transverse $\Jperp$ components of the exchange tensor, and find that they agree with the predictions from the measured single-qubit $g$-tensors. We show that by simply reorienting the magnetic field, we can tune the longitudinal exchange from a conventional positive value through zero, and even induce an effectively negative longitudinal exchange $\Jpara<0$, directly measured for the first time. Furthermore, we demonstrate how the two exchange components are responsible for different types of two-qubit interactions, measuring controlled-phase (CPhase) and SWAP-like oscillations and comparing their evolution frequencies to the spectroscopic measurements. Lastly, by changing the magnetic field orientation and effectively turning off the longitudinal part $\Jpara = 0$, we implement a single-pulse baseband iSWAP gate\cite{ni_diverse_2025}; a proof-of-concept of how engineering the $\Jpara$ vs. $\Jperp$ contributions can enable otherwise-unrealisable two-qubit gates. This work not only provides the first comprehensive characterisation of anisotropic exchange in germanium but also links the $g$-tensor to $J$ engineering, and hence the two-qubit interaction Hamiltonian, paving the way for tunable native two-qubit gates\cite{wu_noiseprotected_2026} and advanced quantum simulations.

\begin{figure*}[htp]
\includegraphics[width=180mm]{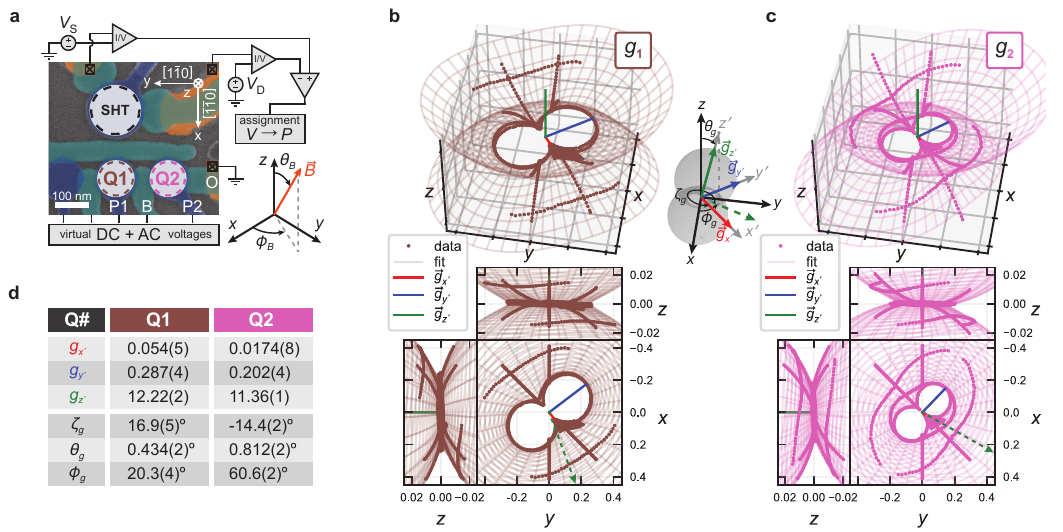}
\caption{\textbf{Device and single-qubit $g$-tensor characterisation}: 
    \textbf{a}, False-coloured SEM image of the device comprising a DQD and SHT. The panel includes a scheme of the control/readout setup, the orientation of the external magnetic field $\B$ and the crystallographic direction in the ($x,y,z$) lab frame.
    \textbf{b} and \textbf{c}, Extracted $g$-tensors of qubit Q1 and Q2 measured at fixed magnetic field magnitude $B=60$~mT plotted in 3D (top panels) and 2D-projections (bottom panels). The red, blue, and green solid lines represent the directions of the three $g$-tensor eigenvectors ($\vec{g}_{x'}$, $\vec{g}_{y'}$, $\vec{g}_{z'}$). The dashed green line in the $xy$-projections represent the tilt direction (projection of $\vec{g}_{z'}$ onto the $xy$-plane).
    \textbf{d}, Eigenvalues ($g_{x'}$, $g_{y'}$, $g_{z'}$) of the $g$-tensor matrix and Euler angles ($\zeta_g$, $\theta_g$, $\phi_g$) describing how the $g$-tensor is rotated in the lab frame.
    }
\labelFig{gs}
\end{figure*}

\section{\labelSec{g_tensor} $g$-tensor characterisation}
The device used in the experiment, shown in \refFig{gs}a, comprises a gate-defined double quantum dot (DQD) fabricated on a planar Ge/SiGe heterostructure\cite{massai_impact_2024}. A neighbouring single-hole transistor (SHT) is capacitively coupled to the DQD and serves as a charge sensor for current-based readout. The DQD is tuned to host two single-hole spin qubits Q1 and Q2, accumulated in the 20-nm strained germanium quantum well (QW) and electrostatically confined under plunger P1 and P2, respectively. The exchange coupling between the qubits is controlled by adjusting the interdot detuning voltage, defined as $\Ve\coloneqq V_{\rm P2}-V_{\rm P1}$, and the interdot barrier voltage $\Vb$. All involved gate voltages are virtualised to correct for cross-capacitances. State preparation and measurement (SPAM) sequences are described in the Methods. The single-hole $g$-tensors are characterised via adiabatic rapid passage (ARP) spectroscopy~\cite{landau_theory_1932, zener_nonadiabatic_1932, wu_population_2011, shafiei_resolving_2013} as a function of the magnetic field orientation at fixed $B =|\B|= 60$~mT. For each $\B$ direction, defined by the azimuth and zenith angles ($\phi_B,\theta_B$), the system is initialised in the $\kdd$ state at the centre of the (1,1) charge region. Here, a chirped microwave pulse on the barrier gate is used to measure the Larmor frequencies $f\qi$ and $f\qii$. The $g$-tensors are then extracted via the Zeeman relation, $hf_{\text{Q}i} = |\mu_{\rm B}\g_{i}\B|$, and presented in \refFig{gs}b,c. Each $g$-tensor is modelled as a diagonal matrix with principal values $g_{x'}<g_{y'}<g_{z'}$, which is then oriented in the laboratory frame by a $zyz$ Euler rotation matrix $R(\zeta_g, \theta_g, \phi_g)$~\cite{hendrickx_sweetspot_2024}. The best-fit parameters are listed in \refFig{gs}d. The equator planes ($x'y'$-planes) of both qubits are tilted by $\theta_g<1^\circ$ with respect to the laboratory $xy$-plane and towards the same quadrant, as visualised by the projections of $\vec{g}_{z'}$ (green dashed arrows) in \refFig{gs}b,c. Furthermore we observe a preferential orientation of the $\vec{g}_{x'}, \vec{g}_{y'}$ eigenvectors for both qubits in the respective $x'y'$-planes, defined by the condition $\phi_g+\zeta_g\simeq 45^\circ$. The current understanding attributes this anisotropy to long-range strain in the QW\cite{seidler_spatial_2025}. The energy splitting between the two qubits, also known as Zeeman energy detuning, arises from their different $g$-factors at a fixed $\B$, and is defined as $\Edg \coloneqq \mu_{\rm B}\dg B = h(f\qi-f\qii)$, shown in \refFig{spectroscopy}a. For Q1, we characterise the drivability, Ramsey and Hahn echo coherence times, respectively $f_{\rm Rabi}/V_{\rm drive}$, $T_2^*$ and $T_2^H$, in the proximity of the $x'y'$-plane, where the hole-Ge nuclei hyperfine coupling is suppressed\cite{hendrickx_sweetspot_2024}. We achieve $T_2^* = 21.3$~$\mu$s in the ergodic limit ($t_{\rm int}=20.9$~min) (\refSuppFig{T2_last_ergodic_fit_45MHz}), which is consistent with the highest reported values for hole spin qubits in Ge\cite{hendrickx_sweetspot_2024, yu_optimising_2026, zeng_highfidelity_2026} (more details in the Supplementary Information \refSec{coherence}).
 % ($\Ve = 15$~mV, $\Vb = 10$~mV)

% \section{\labelSec{g_tensor} Exchange Anisotropy : Theory}
\section{\labelSec{theory} The Origin of Anisotropic Exchange}
The emergence of anisotropic exchange interaction is a direct consequence of spin-orbit effects acting on two coupled spin qubits viewed in their respective eigenframes. Each QD has a unique anisotropic $g$-tensor that encapsulates the overall effect of the QW confinement, electrostatic potential and strain. The interdot coupling is then described by two terms, the spin-conserving and the spin-flip tunnel couplings, with the latter being induced by SOI.
The virtual transfer of a hole to the adjacent dot thus produces an effective anisotropic exchange matrix, that absorbs the site-dependent g-tensors and spin-flip tunnelling. The matrix is composed by an isotropic real term $J_0$, dressed by three composed rotations, written as:
\begin{equation}
    \tensor{J}(\B) = J_0 \R_1(\B)\times\R\SO(-2\theta\SO)\times\R_2^T(\B). 
    \labelEq{J_tensor}
\end{equation}
where $\R_{1,2}$ are the rotations used to diagonalise the single-qubit Zeeman terms and $\R\SO(-2\theta\SO)$ is the counter-clockwise rotation matrix around the unit spin-orbit vector $\nso$ by an angle $-2\theta\SO$, which is the spin-flip angle. Effectively, this complex tensor interaction can be decomposed into two non-equal components responsible for two-qubit dynamics, shown in the energy spectrum in \refFig{spectroscopy}a:

\begin{figure*}[htp!]
\includegraphics[width=180mm, height=95mm]{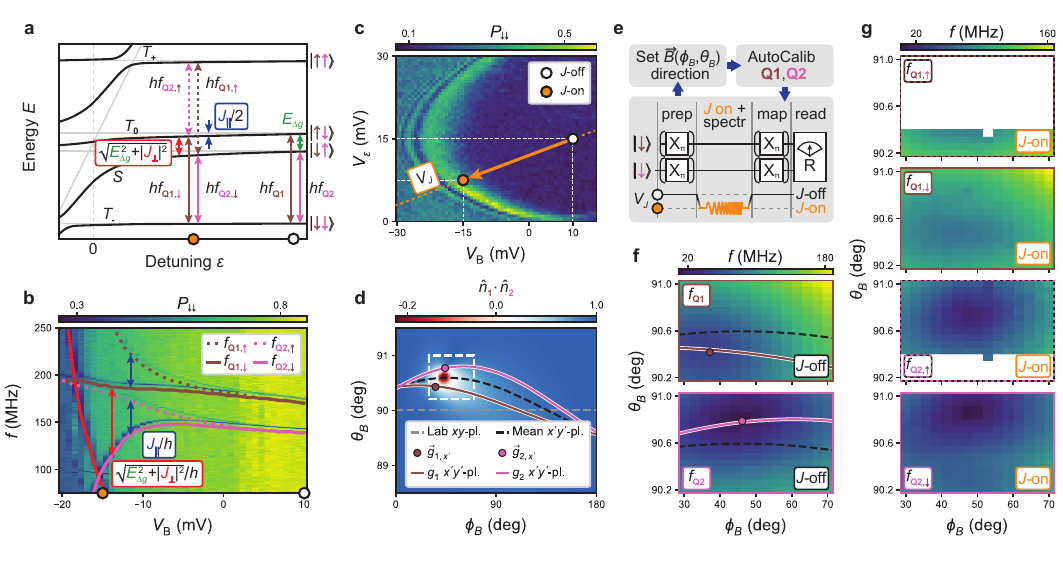}
\caption{\textbf{Spectroscopy of the transition frequencies of the two exchange-coupled qubits}: 
    \textbf{a}, Energy diagram of the two-spin system as a function of the interdot detuning $\varepsilon$ in the (1,1) region. The brown(pink) solid and dashed double-arrows represent the transition frequencies of Q1(Q2) to either $\kdd$ or $\kuu$, for exchange off (white dot) and exchange on (orange dot). The red(blue) arrow represents the energy splitting caused by the transverse $\Jperp$ (longitudinal $\Jpara$) exchange components.
    \textbf{b}, Exchange fans as a function of interdot barrier voltage $\Vb$. The solid and dashed lines indicate the transitions shown in panel \textbf{a}, shifted by 5~MHz for better visibility of the background measurement. The red line indicates the visible transition frequencies relative to the red arrow.
    \textbf{c}, Exchange fingerprint plot\cite{reed_reduced_2016} as a function of the interdot barrier $\Vb$ and detuning $V_{\varepsilon}$ voltages. We plot the singlet probability after a 16-ns decoupled exchange pulse to ($\Vb$,$\Ve$). The composite $V_J$ voltage is used to turn on the exchange.
    \textbf{d}, Simulated scalar product of unit Larmor vectors as a function of the magnetic field angles ($\phi_B,\theta_B$) in the vicinity of the lab $xy$-plane. The unit Larmor vectors are calculated from the fitted $g$-tensors. We plot the minimum and the equator $x'y'$-plane of each qubit's $g$-tensor and their mean $x'y'$-plane.
    \textbf{e}, Spectroscopy measurement protocol used to map the transition frequencies as a function of $\B$ field direction.
    \textbf{f} and \textbf{g}, Measured transition frequencies as a function of the $\B$ field azimuth ($\phi_B$) and zenith ($\theta_B$) in the 2D angular region highlighted in panel \textbf{d} (white dashed square), respectively for exchange off (\textbf{f}) and exchange on (\textbf{g}). The missing data is caused by unreliable single-qubit preparation, due to faulty AutoCalib.
    }
\labelFig{spectroscopy}
\end{figure*}

\paragraph*{\textbf{Longitudinal exchange $\Jpara$} $-$} This term corresponds to an Ising-like $ZZ$ interaction with Hamiltonian $H_{\parallel} = \frac{1}{4}\Jpara\, \sigma_{1}^{z}\sigma_2^{z}$. It shifts each qubit's transition frequency conditional on the state of the other, without inducing state transitions, thereby driving controlled-phase (CPhase) oscillations.

\paragraph*{\textbf{Transverse exchange $\Jperp$} $-$} This term corresponds to a flip-flop $XY$ interaction with Hamiltonian $H_\perp = \frac{1}{2}(\Jperp\, \sigma_1^+ \sigma_2^- + \text{h.c.})$. It couples the two antiparallel states $\kud \leftrightarrow \kdu$, driving iSWAP oscillations.\\

While the exchange anisotropy in silicon holes is dominated by spin-flip tunnelling\cite{geyer_anisotropic_2024}, in germanium holes it is caused mainly by the large $g$-tensor variations\cite{saez-mollejo_exchange_2025}, rather than by the relatively weak spin-flip tunnelling\cite{seidler_spatial_2025} ($\theta\SO\sim10^\circ$). Hence, to first order in $\theta\SO$, $\R\SO(-2\theta\SO)\approx\mathbb{1}$, which leads to:
\begin{subequations}
\labelEq{J_first_order_main}
\begin{align}
\Jpara &= J_0\,\nvi\cdot\nvii \labelEq{J_first_order_para}\\
|\Jperp| &= (J_0+\Jpara)/2 \labelEq{J_first_order_perp}
\end{align}
\end{subequations}
where $\hat{n}_{i}$ is the unit Larmor vector of qubit Q$i$ for a given $\B$, defined as $\hat{n}_i = \g_i\B/|\g_i\B|$. Full derivation in Supplementary Information \refSec{met:Jparallel}.\\

The two-spin dynamics, however, are governed not only by the exchange components but also by the Zeeman energy detuning $\Edg$. Whereas the CPhase evolution is set by $\Jpara$ alone and is directly measurable as $f_{\rm{CPhase}} = |\Jpara|/h$, the dynamics within the antiparallel subspace are jointly determined by $\Jperp$ and $\Edg$: the transverse exchange $\Jperp$ drives the population swap $\kud \leftrightarrow \kdu$, while the detuning $\Edg$ renders it off-resonant, reducing the amplitude of the population transfer. The SWAP-like oscillation frequency is consequently the quadrature sum of these two orthogonal contributions, $f_{\rm{SWAP}} = \left(\Edg^2 + |\Jperp|^2\right)^{1/2}\!\!/h$, and a clean iSWAP requires $\Edg = 0$, a condition we exploit in \refSec{iSWAP}. The SOI in holes thus transforms the exchange into a rich, anisotropic coupling whose two components, together with the Zeeman detuning, govern fundamentally different two-qubit operations.

\begin{figure*}[htp]
\includegraphics[width=180mm, height=96mm]{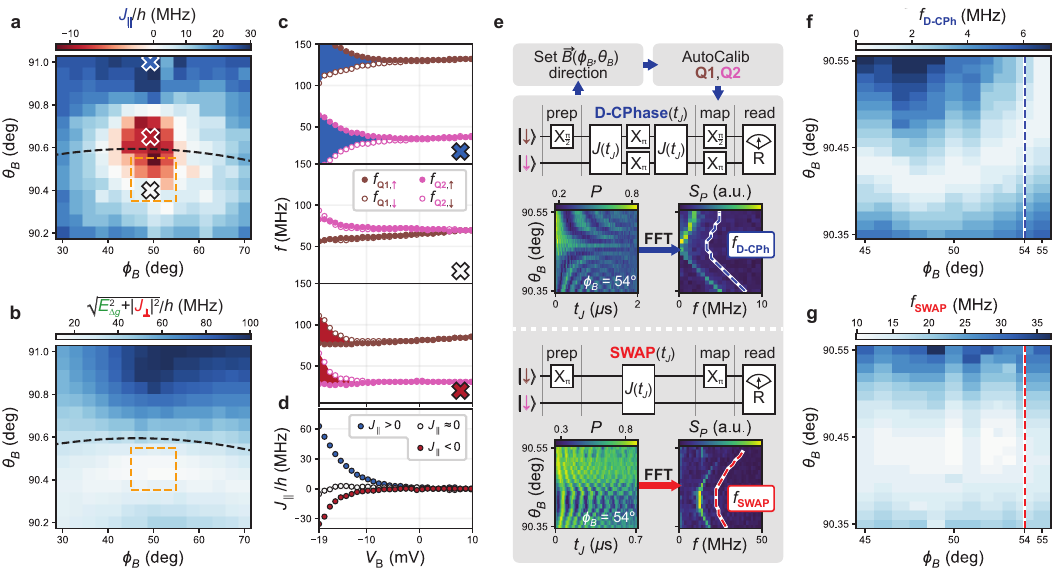}
\caption[Measured anisotropic exchange components and two-qubit gate oscillations]{\textbf{Anisotropic exchange components and two-qubit gate oscillations}: 
    \textbf{a} and \textbf{b}, Longitudinal exchange and quadrature sum of the Zeeman energy detuning and the transverse exchange as a function of magnetic field direction, $(\phi_B, \theta_B)$, in the selected 2D angular region shown in \refFig{spectroscopy}d. The black dashed line represent the mean $x'y'$-plane of the two $g$-tensors. The positive and negative sides of the colour map in \textbf{a} have different scales for better visibility. 
    \textbf{c}, Exchange spin-fans as a function of the interdot barrier voltage, $\Vb$, for three $\B$ directions indicated in \textbf{a} by the blue, white, and red crosses (top, mid, and bottom panel, respectively).
    \textbf{d}, Longitudinal exchange as a function of $\Vb$ extracted from the spin-fans (averaged among the two pairs) shown in \textbf{c}, respecting the colour code for the three $\B$ directions. The different curvatures demonstrate the effective anisotropy of $\Jpara$.
    \textbf{e}, Measurement protocol used to map the D-CPhase and the SWAP-like oscillation frequencies, shown in panels \textbf{f} and \textbf{g}, as a function of the $B$ field direction. The 2D colour-meshes show the measured oscillations (D-CPhase, top inset and SWAP-like, bottom inset) as a function of exchange pulse time $t_J$ and $\theta_B$ for a fixed $\phi_B$ (left) and the respective FFT (right). Dashed lines are the extracted frequencies (shifted for visibility). 
    \textbf{f} and \textbf{g},~D-CPhase and SWAP-like oscillation frequencies as a function of the $B$ field direction. The dependence on ($\phi_B,\theta_B$) mirrors what observed in \textbf{a} and \textbf{b} (orange dashed square). The blue and red dashed lines indicate the angle where the insets in \textbf{e} where measured.
    }
\labelFig{exchange}
\end{figure*}

\section{\labelSec{spectroscopy} Spectroscopic characterisation of Exchange Anisotropy}
We now characterise the longitudinal exchange $\Jpara$, and the transverse coupling term $(\Edg^2 + |\Jperp|^2)^{1/2}$, as a function of the magnetic field direction using ARP spectroscopy.

When the exchange interaction is activated, the qubit frequencies, $f\qi$ and $f\qii$ (previously mapped in \refSec{g_tensor}), split into two non-degenerate pairs: $\{f\qid,f\qiu\}$ and $\{f\qiid,f\qiiu\}$, as schematised in \refFig{spectroscopy}a. This splitting is caused by the longitudinal exchange, $\Jpara$, which can be measured directly from the frequency differences:
$$ \Jpara/h = f\qiu - f\qid = f\qiiu - f\qiid $$
On the other hand, the transverse exchange, $\Jperp$, can only be measured in combination with the Zeeman energy detuning $\Edg$, through the expression:
$$ (\Edg^2 + |\Jperp|^2)^{1/2}/h = |f\qid - f\qiid| = |f\qiu - f\qiiu| $$
To map these four frequencies, we define two operating points in the $(\Ve,\Vb)$ voltage space: a `$J$-off' point for single-qubit manipulation and a `$J$-on' point for two-qubit interaction (see \refFig{spectroscopy}c). We also define the exchange voltage $V_J$ as a combination of $\Ve$ and $\Vb$. The sequence begins by initialising in the $\kdd$ state at the $J$-off point. We then prepare the system in one of the $\{\kdd,\kud,\kdu\}$ states by applying selective single-qubit X$_{\pi}$ gates. Subsequently, we adiabatically ramp to the $J$-on point and apply a broadband ARP pulse to measure the transition frequencies $f\qid$, $f\qiu$, $f\qiid$, and $f\qiiu$. \refFig{spectroscopy}b shows a typical measurement where the two pairs of exchange-split frequencies diverge exponentially as the exchange interaction is increased by lowering $\Vb$, for an arbitrary magnetic field orientation ($\phi_B = 0, \theta_B = 90^\circ$).

\refEq{J_first_order_para} predicts that $\Jpara$ will vary the most when $\B$ lies in the angular region enclosed by the $x'y'$-planes of the two $g$-tensors, as shown in \refFig{spectroscopy}d. Within this region, and particularly near their global minima $\vec{g}_{i,\textit{x'}}$, the large out-of-plane $g$-tensor anisotropy can cause the two Larmor vectors to point in perpendicular or even antiparallel directions. To map the four transition frequencies in this 2D region (white dashed box in \refFig{spectroscopy}d), we adopt the protocol described in \refFig{spectroscopy}e. For each new magnetic field orientation ($\phi_B,\theta_B$), an automated calibration routine (AutoCalib) recalibrates the single-qubit gates at the $J$-off point (Methods and \refSuppFig{autocalib}). ARP spectroscopy is then performed at the $J$-on point, where different transitions become visible depending on the prepared initial state. This entire protocol is repeated to scan $\B$ across the chosen 2D-region. The extracted transition frequencies are plotted as a function of ($\phi_B,\theta_B$) in \refFig{spectroscopy}g. For comparison, \refFig{spectroscopy}f shows the corresponding single-qubit frequencies, $f\qi$ and $f\qii$, measured at the $J$-off point, which are also used for the $g$-tensor fit. Additionally, we note that via spectroscopy, we can map the double-spin-flip frequency $f_{\kdd\leftrightarrow\kuu}$, shown in \refSuppFig{transitions}. From the four measured transition frequencies, we extract $\Jpara/h = \text{mean}(f\qiu - f\qid, f\qiiu - f\qiid)$ and $(\Edg^2 + |\Jperp|^2)^{1/2}/h = \text{mean}(|f\qid - f\qiid|, |f\qiu - f\qiiu|)$. The results are plotted in \refFig{exchange}a and b, respectively. While the quadrature-sum term is always positive and non-zero, dominated by the shape of $|\Edg|$, the longitudinal exchange $\Jpara$ exhibits a dramatic variation across the mapped region. In agreement with the expected dependence of $\nvi\cdot\nvii$, $\Jpara$ transitions from an otherwise mostly isotropic positive value (blue area) to a negative minimum (red area), passing through a distinct circular zero-line (white). To verify this unconventional sign-changing behaviour, we perform separate exchange vs.~$\Vb$ measurements at three distinct $\B$ field directions, marked by the blue, white, and red crosses in \refFig{exchange}a. The results are shown in the three top panels of \refFig{exchange}c. For the direction marked by the blue cross, the exchange interaction behaves as ordinary: the transition frequencies to the $\kdd$ state, $\{f\qid,f\qiid\}$, bend downwards, while those to the $\kuu$ state, $\{f\qiu,f\qiiu\}$, bend upwards. At the white-cross direction, however, the frequencies remain degenerate and do not split as $\Vb$ is decreased. Remarkably, for the red-cross direction, the frequency branches diverge again, but with the opposite sign, providing a clear signature of effective negative exchange. This is the first direct measurement demonstrating that the longitudinal exchange term $\Jpara$ can be controllably tuned to be positive, null, or negative, all while maintaining its characteristic exponential dependence on $\Vb$, as shown in \refFig{exchange}d.

\section{\labelSec{tqgate} Two-Qubit Interactions}
Next, we demonstrate the direct connection between the exchange components and their respective two-qubit gate interactions. By diabatically turning on and off the exchange interaction via baseband voltage pulses on $V_J$, we can observe CPhase and SWAP-like oscillations as a function of the pulse duration, $t_J$. To map these oscillations versus the magnetic field direction, we adopt the protocol described in \refFig{exchange}e: for each orientation $\B(\phi_B,\theta_B)$, we run the AutoCalib routine before the gate measurements. The conditional phase driving the CPhase gate is measured by preparing Q1 in $\ket{-i} = (\kd-i\ku)/\sqrt{2}$ and applying a decoupled controlled-phase (D-CPhase) gate\cite{watson_programmable_2018}, composed by two identical exchange pulses (fixed $V_J$, varying $t_J$) interleaved by an X$_\pi$ on each qubit, which echoes out the additional phase accumulated due to the $g$-tensor voltage dependence $\text{d}g/\text{d}V_J$. The SWAP-like oscillations between antiparallel states are measured by preparing $\kud$ and applying a single exchange pulse of varying duration $t_J$. In both cases we read out $P(t_J)$ after mapping back to the measurement basis as described in the Methods. From the resulting probability oscillations, we extract the characteristic frequencies, $f_{\rm D-CPh}$ and $f_{\rm SWAP}$, using a Fast Fourier Transform (FFT). The left-side insets in \refFig{exchange}e show examples of the measured probability oscillations as a function of $t_J$ and $\theta_B$ for a fixed $\phi_B$, while the right-side insets show the corresponding FFTs. The extracted frequency traces for $f_{\rm D-CPh}$ (dashed blue line) and $f_{\rm SWAP}$ (dashed red line) are clearly visible. Following this protocol, we map $f_{\rm D-CPh}$ and $f_{\rm SWAP}$ across the reduced 2D angular region indicated by the orange dashed box in \refFig{exchange}a,b and plot the results in \refFig{exchange}f. The extracted oscillation frequencies agree with the spectroscopic measurements through the relations $f_{\rm{D-CPh}} = |\Jpara|/h$ and $f_{\rm{SWAP}} = (\Edg^2 + |\Jperp|^2)^{1/2}/h$. We note, however, that $f_{\rm{D-CPh}}$ is sensitive only to the magnitude of the accumulated conditional phase and not to its sign, which is not resolved by this measurement. This is visually confirmed in the upper-left inset in \refFig{exchange}e, where the oscillation pattern shows no evident change as it crosses the zero-frequency line at $\theta_B\sim90.47^\circ$. With this caveat, comparing $|\Jpara|$ from the two methods provides an independent confirmation of the spectroscopic result and a more precise localisation of the $\Jpara = 0$ line.

\begin{figure}[htp]
\includegraphics[width=90mm, height=97mm]{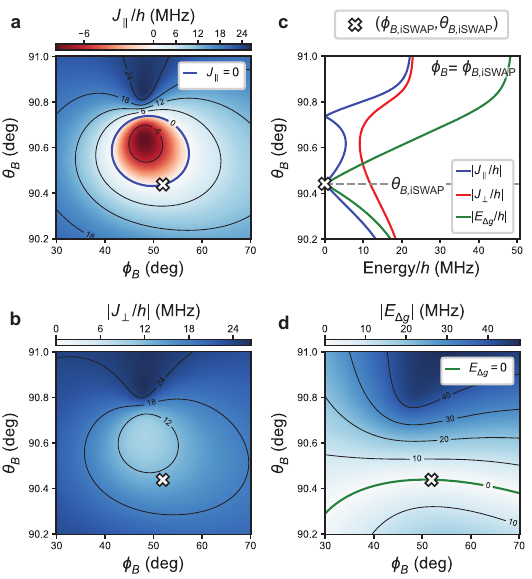}
\caption[Simulated anisotropic exchange components from best fitting parameters]{\textbf{Simulated anisotropic exchange components from best fitting parameters}: 
    \textbf{a},~\textbf{b},~ and \textbf{d},~Separate simulations of the longitudinal exchange $\Jpara$, the perpendicular exchange $\Jperp$ and the Zeeman energy detuning $\Edg$, respectively, as a function of the $\B$ field directions ($\phi_B,\theta_B$) using the best fitting parameters reported in \refSuppTab{fit_param}. The white cross indicates the magnetic field direction ($\phi_{B,\rm iSWAP},\theta_{B,\rm iSWAP}$) where the baseband iSWAP gate is implemented. 
    \textbf{c},~Modules of the exchange components and Zeeman energy detuning at fixed $\phi_B = \phi_{B,\rm iSWAP}$ showing the simultaneous zeros ($\Jpara=\Edg=0$) at $\theta_B = \theta_{B,\rm iSWAP}$.
    }
\labelFig{simulations}
\end{figure}

\begin{figure*}[htp]
\includegraphics[width=180mm]{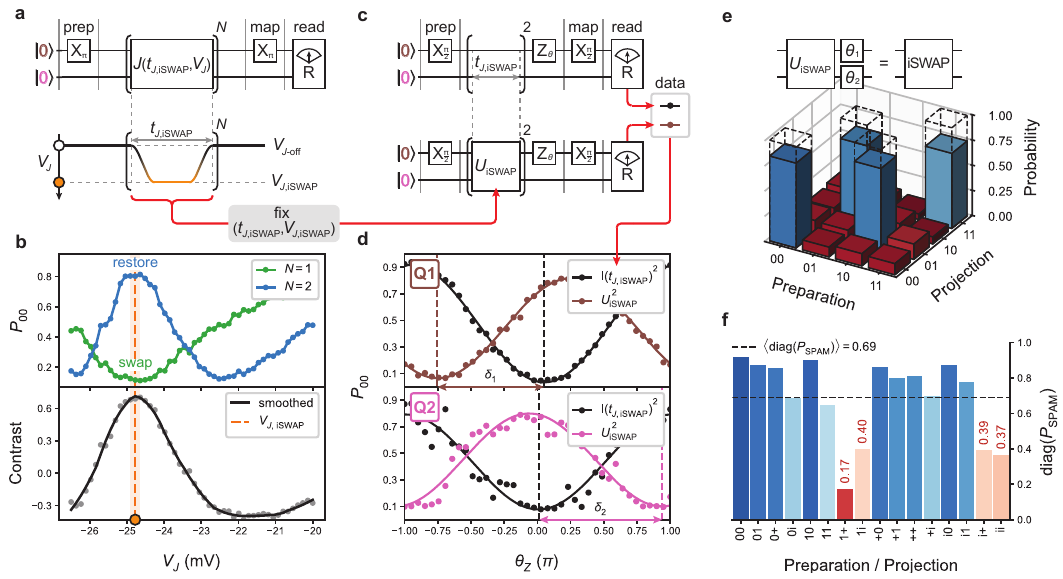}
\caption[Calibration and process tomography of the baseband iSWAP]{\textbf{Calibration and process tomography of the baseband iSWAP}: 
    \textbf{a}, Gate sequence for exchange-amplitude calibration: after preparing $|{\uparrow}{\downarrow}\rangle$, $N$ identical exchange pulses (duration $t_{J,\mathrm{iSWAP}}$, amplitude $V_J$, Tukey-windowed) are applied before mapping and readout.
    \textbf{b}, Top: Measured probability $P_{00}$ versus exchange amplitude $V_J$ for $N=1$ (green) and $N=2$ (blue); a calibrated single pulse swaps the antiparallel spin populations while a doubled pulse restores them. Bottom: their contrast $P_{00}^{N=2}-P_{00}^{N=1}$ (grey) and smoothed envelope (black); the peak (orange) sets $V_{J,\mathrm{iSWAP}}$.
    \textbf{c}, Gate sequence for single-qubit phase calibration (Q1): $U_{\mathrm{iSWAP}}^2$ (or an idle reference of matched duration) is followed by a swept phase gate $Z(\theta_Z)$, applied independently to each qubit.
    \textbf{d} $P_{00}$ versus $\theta_Z$ for Q1 (top, brown) and Q2 (bottom, pink), for the idle reference (black) and $U_{\mathrm{iSWAP}}^2$ (coloured). The fringe offsets $\delta_i$ (coloured double-arrows) are used to extract the single-qubit phase corrections $\theta_i$.
    \textbf{e}, Top: the calibrated iSWAP is the exchange pulse $U_{\mathrm{iSWAP}}$ plus single-qubit virtual phase corrections, $Z(\theta_1)$ and $Z(\theta_2)$. Bottom: partial QPT projected on the computational basis; bars are measured probabilities (no SPAM correction), black dashed wireframes indicate the ideal iSWAP.
    \textbf{f}, Diagonal of the measured 16-state SPAM matrix; dashed line marks the mean. Bars are coloured according to the probability maps in \refSuppFig{FullQPT}.
}
\labelFig{iSWAP}
\end{figure*}

\section{\labelSec{iSWAP} Baseband \MakeLowercase{i}SWAP via Hamiltonian engineering}

Using the measured two-dimensional maps of the transition frequencies, we extract the spin-orbit angle $\theta\SO$ and vector $\hat n\SO$, alongside the parameters describing the $g$-tensors in the coupled regime (see Supplementary Information \refSec{met:fitting} and \refSuppTab{fit_param} for fitting details). This allows us to separately simulate and visualise $\Jpara$, $\Jperp$, and $\Edg$, within the angular region of interest. The results are displayed in \refFig{simulations}. $\Jpara$ closely matches expectations from the measurements (\refFig{exchange}a,f), exhibiting a zero-line enclosing a negative region. $\Jperp$ shows a dependence on $\B$ similar to $\Jpara$, but never vanishes, in agreement with \refEq{J_first_order_perp}. $\Edg$ has a zero line where the two $g$-tensors intersect.

The anisotropy with respect to $\B$, combined with the disparity between the exchange components, has significant implications for two-qubit operations.
For instance, inadvertently operating at $\Jpara\approx0$ vanishes conditional phases and precludes a native CPhase gate. 
However, while it can degrade standard control schemes, it also grants access to gates unattainable with isotropic exchange. As shown in the cross-section in \refFig{simulations}c, our $g$-tensor tuning yields a specific magnetic field direction (indicated by the dashed grey line and white cross) where $\Jpara = \Edg = 0$ simultaneously, while $\Jperp\approx J_0/2$. By aligning the magnetic field along this direction, we can effectively engineer the Hamiltonian to a pure transverse ($XY$) exchange interaction, taking the form in the $\{\kuu,\kdu,\kud,\kdd\}$ basis:
\begin{equation}
    H = 
    \begin{pmatrix}
    \bar{E}_{\rm Z} & 0 & 0 & 0 \\
    0 & 0 & \Jperp/2 & 0 \\
    0 & \Jperp^*/2 & 0 & 0 \\
    0 & 0 & 0 & -\bar{E}_{\rm Z}
    \end{pmatrix}.
    \labelEq{exchange_matrix}
\end{equation}
We note that, due to $\text{d}g_{i}/\text{dV}_{J}\neq 0$, turning off the exchange yields $\Edg\neq0$, preserving single-qubit addressability.
In this configuration, applying a single baseband exchange pulse for a duration $t_{J,\rm iSWAP} = h(2\Jperp)^{-1}$ realises an iSWAP gate. 

% Realising the iSWAP requires calibrating the amplitude of the baseband exchange pulse, which sets the swap angle, and the residual single-qubit $Z$ phases corrections. We calibrate the amplitude by preparing $\kud$ and applying $N$ consecutive exchange pulses of fixed duration $t_{J,\mathrm{iSWAP}}=56$~ns before reading out the return population $P_{00}$ as a function of the pulse amplitude $V_J$ (\refFig{iSWAP}a). At a correctly tuned amplitude a single pulse fully swaps the two qubits while two pulses restore the initial state, so the contrast between the $N=1$ and $N=2$ traces is maximised. Its peak defines the working point $V_{J,\mathrm{iSWAP}}$, as shown in \refFig{iSWAP}b. Because a single iSWAP exchanges the two qubits, scrambling any single-qubit phase reference, we measure the accumulated $Z$ phase using a doubled gate, $\mathrm{iSWAP}^2$, which returns each qubit to itself (\refFig{iSWAP}c). Sweeping the phase of a virtual gate $Z(\theta_Z)$ before readout and comparing the $\mathrm{iSWAP}^2$ interference fringe against an idle reference gives the total phase over two pulses; halving the offset yields the per-iSWAP correction $\theta_i$ applied to each qubit (\refFig{iSWAP}d, full details on the exchange pulse shape in \refSuppFig{iSWAP_pulse}). The calibrated iSWAP is the exchange pulse $U_{\mathrm{iSWAP}}$ followed by these two single-qubit $Z$ rotations, as schematised in \refFig{iSWAP}e, top.

Realising the iSWAP requires calibrating the amplitude of the baseband exchange pulse, which sets the swap angle, and the residual single-qubit $Z$ phase corrections. We calibrate the amplitude by preparing $\kud$ and applying $N$ consecutive exchange pulses of fixed duration $t_{J,\mathrm{iSWAP}}=56$~ns before reading out the return population $P_{00}\coloneqq P_{\downarrow\downarrow}$ as a function of the pulse amplitude $V_J$ (\refFig{iSWAP}a). At a correctly tuned amplitude a single pulse fully swaps the two qubits while two pulses restore the initial state, so the contrast between the $N=1$ and $N=2$ traces is maximised. Its peak defines the working point $V_{J,\mathrm{iSWAP}}$, as shown in \refFig{iSWAP}b (details on the pulse shape in \refSuppFig{iSWAP_pulse}). Because a single iSWAP exchanges the two qubits, mixing-up single-qubit phase references, we measure the accumulated $Z$ phase using a doubled gate, $\mathrm{iSWAP}^2$, which returns each qubit to itself (\refFig{iSWAP}c). An ideal  $\mathrm{iSWAP}^2 = Z\otimes Z$, which imprints a deterministic $\pi$ phase on each qubit, on top of the accumulated spurious single-qubit phases. Sweeping the phase of a virtual gate $Z(\theta_Z)$ before readout and comparing the $\mathrm{iSWAP}^2$ interference fringe against an idle reference returns a total phase offset $\delta_i$; the per-iSWAP correction applied to each qubit is then $\theta_i = \pi/2 - \delta_i/2$, where the $\pi/2$ retains the ideal $Z \otimes Z$ phase and $\delta_i/2$ cancels the spurious contribution (\refFig{iSWAP}d). The calibrated iSWAP is the exchange pulse $U_{\mathrm{iSWAP}}$ followed by these two single-qubit $Z$ rotations, as schematised in \refFig{iSWAP}e, top.

We characterise the calibrated gate by quantum process tomography (QPT) in an informationally-complete 16-state basis built from all tensor products of $\{\ket{0}\coloneqq \kd, \ket{1}\coloneqq \ku, \ket{+}, \ket{i}\}$. \refFig{iSWAP}e shows the partial process map projected onto the computational basis $\{|00\rangle, |01\rangle, |10\rangle, |11\rangle\}$, which directly reveals the iSWAP signature: the parallel states $|00\rangle$ and $|11\rangle$ are preserved, while the antiparallel states $|01\rangle$ and $|10\rangle$ are exchanged. From the full 16×16 dataset, shown in \refSuppFig{FullQPT}, we extract a process fidelity $F_\mathrm{QPT} = 60\,\%$. This value is dominated by SPAM errors rather than intrinsic gate error: the diagonal of the independently-measured SPAM matrix, reported in \refFig{iSWAP}f, has a mean of $\langle\mathrm{diag}(P_\mathrm{SPAM})\rangle = 0.69$, which caps the achievable $F_\mathrm{QPT}$ near $70\,\%$ even for a noiseless gate. Modelling each SPAM operation as a depolarising channel\cite{merkel_selfconsistent_2013} gives $F_\mathrm{QPT} \approx F_\mathrm{gate}\,\langle\mathrm{diag}(P_\mathrm{SPAM})\rangle$, yielding an estimated underlying gate fidelity $F_\mathrm{iSWAP} \approx 87\,\%$ (Supplementary Information \refSec{met:QPT}). The lowest SPAM diagonal values correspond to the states requiring concatenated rotations on both qubits, consistent with the reduced $T_2^*$ at the iSWAP field point, where $J_\parallel = E_{\Delta g} = 0$ forces the magnetic field out of the $g$-tensor equatorial planes. A more rigorous fidelity estimation via randomised benchmarking\cite{magesan_characterizing_2012} is left to follow-up work.

\section{\labelSec{conc} Conclusions}

In conclusion, we have shown that for hole spin qubits in a strained-germanium quantum well, the exchange interaction exhibits a strong anisotropy in magnetic field direction, set by the shape and relative tilt of the single-qubit $g$-tensors. Mapping the $g$-tensors, we measured a Ramsey coherence time of $T_2^* = 21.3~\mu$s in the $g$-tensor equatorial plane in the ergodic limit, among the highest reported for hole spin qubits. Then, via transition-frequency spectroscopy under full vector control of $\B$, we resolved the exchange into longitudinal ($\Jpara$) and transverse ($\Jperp$) components, which follow $\B$ similarly but differ in magnitude. The anisotropy is strongest within the $g$-tensor equatorial planes, near the aligned $\vec{g}_{x'}$ eigenvectors: there $\Jperp$ remains positive while $\Jpara$ sweeps from its positive isotropic value through zero to negative, defining a circular zero-line in $(\phi_B,\theta_B)$. To our knowledge, this is the first direct measurement of effectively negative longitudinal exchange in spin qubits. By investigating two-qubit oscillations driven by baseband exchange pulses, we linked the CPhase and iSWAP frequencies directly to $\Jpara$ and $\Jperp$, establishing the field direction as a continuous control knob over the interaction Hamiltonian. Finally, tuning $\B$ to $\Jpara = \Edg = 0$ with $\Jperp \neq 0$, we reduced the interaction to a pure transverse ($XY$) form and realised an iSWAP with a single $56$-ns baseband pulse, unattainable with isotropic exchange. Process tomography yields $F_\mathrm{iSWAP} \approx 87\%$, limited by single-qubit coherence and readout at an iSWAP point that, for our $g$-tensor configuration, lies outside the hyperfine-insensitive planes.

We note that, both the $\Jpara = 0$ and $\Edg = 0$ conditions are fixed by the relative geometry of the single-qubit $g$-tensors, and are therefore neither guaranteed to exist nor to coincide for an arbitrary pair of dots. In an array with uncontrolled $g$-tensor tilts, no single global field direction will optimise coherence for every qubit, and individual pairs may cross $\Jpara = 0$ unintentionally, suppressing conditional-phase gates where they are required. Characterising and ultimately engineering $g$-tensor geometry is thus a prerequisite for scaling rather than an optional refinement. Tailoring strain and confinement across an array through targeted heterostructure development could bring long coherence, baseband single- and two-qubit gates (via hopping and anisotropic exchange, respectively), and motional-narrowed shuttling into simultaneous operation, positioning germanium as a uniquely versatile platform for scalable quantum hardware.

%TC:ignore
\section{Methods}
\subsection*{\labelSec{met:fab} Device}
The device used in this experiment is a sub-portion of a linear 6-QD array, similar the ones used in Ref.~\cite{tsoukalas_dressed_2026, tsoukalas_resonant_2025, kelly_identifying_2025}, where more details on the fabrication process can be found. Details about the Ge/SiGe heterostructure can be found in Ref.~\cite{bedell_lowtemperature_2020, massai_impact_2024, seidler_spatial_2025}.
In particular, we utilise the rightmost DQD plus the SHT located above, shown in \refFig{gs}a. A colour legend of the false-coloured SEM image can be found in \refTab{mat_legend}. The ohmic contact on the right of QD2 is used for loading/unloading of holes.
\begin{table}[ht]
    \centering
    \renewcommand{\arraystretch}{1.3} 

    % Define vertical lines in the preamble: |c|c|c|
    \begin{tabular}{|c|c|c|} 
        \hline

        \textbf{Colour} & \textbf{Material} & \textbf{Scope} \\
        \hline
        gray & SiO$_2$   & gate oxide \\
        orange & PtSiGe  & ohmic contacts \\
        green & Ti/Pd (5+15~nm) & 1$^{\rm st}$ gate layer \\
        blue & Ti/Pd (5+15~nm)  & 2$^{\rm nd}$ gate layer\\ 
        \hline
    \end{tabular}
    \caption{Colour legend of the false-coloured SEM image in \refFig{gs}a.}
\labelTab{mat_legend}
\end{table}

% \begin{itemize}
%     \item gray - SiO$_2$, gate oxide;
%     \item orange - PtSiGe, ohmic contacts;
%     \item green - Ti/Pd ($5+15$~nm), 1$^{st}$ gate layer;
%     \item blue - Ti/Pd ($5+15$~nm), 2$^{nd}$ gate layer.
% \end{itemize}

\subsection*{\labelSec{met:setup} Experimental setup}
The sample, mounted on a QDevil QBoard circuit board, is loaded in a Bluefors LD400 dilution refrigerator and cooled down to a base temperature of $T \approx 15$ mK. The dc gate voltages are applied using a QDevil QDAC 2 through twisted-pair wiring and filtered using a QDevil QFilter at the millikelvin stage of our fridge. In addition, part of the gates are connected to coaxial lines through on-PCB bias tees. All coaxial lines are attenuated by a total of 23~dB. The ac gate voltages are applied using a Quantum Machines OPX+, that provides both baseband and oscillating signals up to $f=350$~MHz. A representative output trace spanning initialisation, readout, and control, including the baseband exchange pulse and the single-qubit drives, is shown in \refSuppFig{iSWAP_pulse}a. The double quantum dot charge occupation is sensed by the single hole transistors placed above. We apply a dc source-drain bias of $V_{\rm SD} = 250$ $\rm{\mu V}$ and measure the differential current $I_{\rm SD}$ using a pair of BasPI SP983c IV-converters ($\rm gain=10^7$, $f_{\rm cut-off} = 100$~kHz) and a in-house-made differential amplifier ($\rm gain=1$, $Z_{\rm out} = 50~\Omega$, $V_{\rm clip} =\pm 2$~V) connected to the analog input of the OPX+. The rf signals used for Rabi-driving the qubits are applied onto single real gates (P1, P2 or B). The driven real gates are varied to maximise the qubit drivability at different magnetic field orientations. See \refFig{gs}a for a scheme of the connectivity in the setup. The magnetic field $\B = (B_x, B_y, B_z)$ is applied by an American Magnetics three-axis magnet with a maximum field of 1/1/6 Tesla in the $x/y/z$ direction and a high-stability option on all coils. Small common rotations of Q1's and Q2's $g$-tensors may occur due to imperfect planar mounting of the sample. Finally, we note that our magnet coils typically have sub-millitesla offsets due to hysteresis, corrected via the $B$-field offset calibration described in the Methods.

\subsection*{\labelSec{met:B_calib} Magnetic-field offset calibration}
The coils of the three-axis magnet are subject to hysteresis caused by magnetic flux trapping, resulting in millitesla-scale offsets. Due to the large out-of-plane anisotropy of the qubit's $g$-tensor, it is particularly important to correct offsets in $B_z$. 
For example, for a typical $g$-tensor with $g_{\rm diag} = (0.05,0.2,11)$, a $B_{z, \rm offset} = 1$~mT already results in a $f_{\text{Q}i,\rm offset}\approx150$~MHz in the Larmor frequency of qubit Q$i$, while a $B_{x,y, \rm offset} = 1$~mT only causes $f_{\text{Q}i,\rm offset}<3$~MHz. These offsets are altered only with relatively high fields ($\sim1$~T for $B_x, B_y$ and ~100mT for $B_z$). Therefore, after the calibration, we find the correction to hold as long as the magnetic field magnitude is maintained within the mentioned limits. To measure and correct $B_{z, \rm offset}$, we use a single-qubit's frequency and the intrinsic symmetry of the $g$-tensor. We perform two spectroscopic measurements --- indexed with "$+$" and "$-$" --- of the qubit frequency while sweeping $B_z$ around zero, with a fixed $B_x^{\pm}=\pm 20$~mT, as shown in \refSuppFig{calibB_field}a,b. The average between the signals read out during the two measurements is shown in \refSuppFig{calibB_field}c. For each measurement with fixed $B_x^{\pm}$ value, the qubit frequency reaches a minimum $f_{\rm Qi}^{\pm}$ at $B_z^{\pm}$. While the difference $B_{z,\rm diff} = |B_z^{+} - B_z^{-}|$ is an indication of the $g$-tensor tilt, the mean $B_{z,\rm mean} = (B_z^{+} + B_z^{-})/2$ represents the offset $B_{z, \rm offset}$ to be corrected. Additionally, $f_{\rm Qi, diff} = |f_{\rm Qi}^+ - f_{\rm Qi}^-| \neq 0$ indicates the presence of a small $B_{x,\rm offset}$, neglected here.

\subsection*{\labelSec{met:SPAM} State preparation and measurement}
% State preparation and measurement (SPAM) are performed similarly to Ref.\cite{kelly_identifying_2025}. To prepare the $\kdd$ state, we adiabatically ramp the detuning voltage $\Ve$ from the $S(0,2)$ to the (1,1) charge configuration. To read out, we invert this ramp, mapping the $\kdd$ state back to $S(0,2)$ while the states $\{\kud,\kdu,\kuu\}$ remain in (1,1), thus achieving spin-to-charge conversion. During the ramps, a negative pulse on $V_{\rm B}$ increases the interdot tunnel coupling $t_c$, which allows for shorter ramp times $t_{\rm ramp\:in/out}$ and higher SPAM fidelity. Via this protocol, however, it is not possible to directly distinguish between the three blockaded states, hence, in order to study an arbitrary state $\ket{\xi\gamma}$, we always prepare from (and map back to) $\kdd$ via single-qubit gates. The initialisation in $\kdd$ utilises the super-slow adiabatic passage (super-SAP) through the spin-orbit gap $\Delta\SO$ ($ST_-$ anti-crossing) which is known to be anisotropic in germanium~\cite{kelly_identifying_2025, seidler_spatial_2025}. Therefore, the ramp durations $t_{\rm ramp\:in/out}$ are readjusted each time the magnetic field vector is modified, varying within $0.2 - 3$~$\mu$s. By tuning asymmetric reservoir-dot tunnel rates, we implement a double-latched readout technique, similar to Ref.~\cite{kelly_identifying_2025}.

State preparation and measurement (SPAM) are performed similarly to Ref.~\cite{kelly_identifying_2025}. To prepare the $\kdd$ state, we adiabatically ramp the detuning voltage $\Ve$ from the $S(0,2)$ to the $(1,1)$ charge configuration; to read out, we invert this ramp, mapping $\kdd$ back to $S(0,2)$ while the states $\{\kud,\kdu,\kuu\}$ remain in $(1,1)$, thereby achieving spin-to-charge conversion. During the ramps, a negative pulse on $V_{\rm B}$ increases the interdot tunnel coupling $t_c$, allowing for shorter ramp times $t_{\rm ramp\:in/out}$ and higher SPAM fidelity. The initialisation into $\kdd$ relies on the super-slow adiabatic passage (super-SAP) through the spin-orbit gap $\Delta\SO$ (the $ST_-$ anticrossing), which is anisotropic in germanium~\cite{kelly_identifying_2025, seidler_spatial_2025}; the ramp durations $t_{\rm ramp\:in/out}$ are therefore readjusted whenever the magnetic field vector is changed, varying within $0.2 - 3$~$\mu$s. This protocol cannot directly distinguish between the three blockaded states $\{\kud,\kdu,\kuu\}$; to access an arbitrary state $\ket{\xi\gamma}$, we therefore always prepare from, and map back to, $\kdd$ using single-qubit gates, effectively always measuring $P_{\downarrow\downarrow}$. Finally, by tuning asymmetric reservoir-dot tunnel rates, we implement a double-latched readout technique, again following Ref.~\cite{kelly_identifying_2025}.

\subsection*{\labelSec{met:AutoCalib} Automatic single-qubit calibration - AutoCalib}
An automated three-step calibration routine (AutoCalib) is executed on each qubit whenever the magnetic field is changed (in magnitude and/or direction) to a new vector, $\B_{\rm new}$. The first step determines the Larmor frequency $f_{\text{Q}i}$. After initialising the qubit to $\ket{\downarrow}$, we apply an X$_{\pi}$ sweeping its frequency around an estimate $f_{\text{Q}i,0} = \mu_B |\g_{i} \B_{\rm new}|/h$, derived from the $g$-tensor characterisation. Fitting a Gaussian to the measured return probability returns $f_{\text{Q}i}$. Next, the gate amplitudes are calibrated. The X$_{\pi}$ gate amplitude, $V_{\pi}$, is found using error-amplifying sequences of $2n$ and $2n+1$ pulses, with a fixed $t_{\pi} = 456$~ns. The X$_{\pi/2}$ gate amplitude, $V_{\pi/2}$, is determined similarly, with sequences of $4n+1$ and $4n+3$ pulses. A simultaneous sinusoidal fit to the measured state probabilities for each gate determines its optimal amplitude. The AutoCalib procedure, inspired by Ref.\cite{wu_simultaneous_2025}, is shown in \refSuppFig{autocalib}.

\section{Data availability}
All data underlying this study will be made available in a Zenodo repository upon publication.

\section{Acknowledgements}
We acknowledge the staff of the Binnig and Rohrer Nanotechnology Center for their contributions to the sample fabrication and thank all members of the IBM Research Europe - Zurich spin qubit team for useful discussions. We thank M Rimbach-Russ and S Bosco for insightful discussions. We acknowledge funding by NCCR SPIN, a National Centre of Competence in Research, funded by the Swiss National Science Foundation (grants 51NF40-180604 and 51NF40-225153). LM and AF acknowledge support from the Swiss National Science Foundation (grant no. 200021-188752).

\section{Author contributions}
LM performed the experiment and data analysis with contributions from EGK and MA; LM, BH and PHC performed the simulations and developed the theoretical model; FJS and MM fabricated the sample, designed by KT and LS; LM, IS, EGK, SP, GS, and PHC contributed to the development of the experimental setup and measurement software; SWB grew the heterostructures; AO, MPV, UvL and AF contributed to discussions and feedback; LM wrote the manuscript with contributions from BH and input from all authors; PHC supervised the project.

\section{Competing Interests}
The authors declare no competing interests.

% % \bibliographystyle{naturemag_edit} %coincise, no hyperlinks, Nature style
% \bibliographystyle{prb_edit} %long but with hyperlinks, PRB style
% \bibliographystyle{naturemag_doi_arxiv} %coincise, with organised hyperlinks, Nature style, arxiv as well
\bibliography{SpinLibBetterBibTex}

\section{Supplementary Information}
Supplementary information are available with this paper.

\clearpage
% \beginsupplement

\title{Supplementary Information:\\Engineering two-qubit gates via anisotropic exchange in germanium spin qubits}

\maketitle
\renewcommand{\figurename}{Supplementary Figure}
\renewcommand{\tablename}{Supplementary Table}
\onecolumngrid

\section{\labelSec{coherence} Coherence in the \MakeLowercase{\textit{g}}-tensor equator plane}
We investigate the single-qubit properties of Q1 with the magnetic field vector aligned near the $g$-tensor equatorial plane, specifically in the vicinity of the eigenvectors $\vec{g}_{x'}$ and $\vec{g}_{y'}$. To do so, we fix the azimuthal angle $\phi_B$ to lay in the $\vec{g}_{x'}$ and $\vec{g}_{y'}$ directions and sweep the zenith angle $\theta_B$ symmetrically across the $x'y'$-plane. During this sweep, the magnetic field magnitude is adjusted to maintain a constant Larmor frequency $f_{\rm Q1} \approx 100$~MHz. In this configuration, we measure the drivability $f_{\rm Rabi}/V_{\rm drive}$, as well as the Ramsey and Hahn echo coherence times, $T_2^*$ and $T_2^H$, respectively.

The results, plotted in \refSuppFig{T2_vs_theta}, show that both drivability and coherence times exhibit a maximum in the $g$-tensor $x'y'$-plane, consistent with observations in Ref.\cite{hendrickx_sweetspot_2024}.
The drivability is maximised due to the sharp profile of the $g$-tensor anisotropy and the resulting high voltage susceptibility.
Coherence is maximised due to the suppression of the hyperfine coupling to the $^{73}$Ge nuclear spin bath. Spin population revivals (orange dots in \refSuppFig{T2_vs_theta}c, top panel) occur at times proportional to $(\gamma_\text{Ge-73}B)^{-1}$, where $\gamma_\text{Ge-73}$ is the $^{73}$Ge gyromagnetic ratio. 

We now focus on the highest-coherence sweet spot, aligning the magnetic field along $\vec{g}_{y'}$. We measure free induction decay traces via Ramsey experiments, continuously recording for over 20 minutes, and fit each trace to extract $T_2^*$. Furthermore, we fit the cumulative average and plot the results in \refSuppFig{T2_ergodic}a. We find a $T_2^* = 21.3$~$\mu$s in the ergodic limit ($t_{\rm int} = 20.9$~min) at $f_{\rm Q1} = 46.3$~MHz ($B = 9.7$~mT), which is consistent with the highest reported values for hole spin qubits in Ge\cite{hendrickx_sweetspot_2024, yu_optimising_2026, zeng_highfidelity_2026} (time trace shown in \refSuppFig{T2_last_ergodic_fit_45MHz}). By varying the magnetic field, we measure the dependence of the ergodic $T_2^*$ on the qubit Larmor frequency $f_{\rm Q1}$. The results are shown in \refSuppFig{T2_ergodic}b. Consistent with measurements in Ref.\cite{hendrickx_sweetspot_2024}, for large $B$ we observe $T_2^*\propto f_{\rm Q1}^{-1}$, indicating charge-noise limited coherence. At low $B$, the finite spread of precession frequencies within the nuclear spin ensemble limits qubit coherence, causing a decrease in $T_2^*$\cite{bluhm_dephasing_2011}. This indicates that, even at the hyperfine-noise sweet spot, coherence remains limited by coupling to the nuclear spin bath.

\section{\labelSec{met:Jparallel} Parameters of the exchange matrix in vector-algebraic form}

In this section we derive the relevant parameters of the exchange matrix in terms of basis-invariant quantities, namely the $g$-tensors, the magnetic field, and the spin-orbit vector. To this we start from the effective low-energy 2-qubit Hamiltonian in the (1,1) charge sector expressed in terms of the aforementioned tensor quantities
\begin{equation}
    H = \frac{\muB}{2} |\gi\B| \nvi \Pauliveci  + \frac{\muB}{2} |\gii\B| \nvii \Paulivecii + \frac{J_0}{4}\Pauliveci \R\SO \Paulivecii\,
    \label{eq:Hlab}
\end{equation}
where $\R\SO$ is the rotation of a $-2\theta\SO$ angle around the spin-orbit vector $\nso(\alpha_{n_{SO}},\beta_{n_{SO}})$ with $\{\alpha_{n_{SO}},\beta_{n_{SO}}\}$ as azimuth and zenith angles, respectively;  and $\nvi = \gi\B/|\gi\B|$ (similarly for $\nvii$). Furthermore $\Pauliveci  = (\sigma_x \hat{e}_x + \sigma_y \hat{e}_y + \sigma_z \hat{e}_z)\otimes \mathbb{1}$ and $\Paulivecii  = \mathbb{1}\otimes(\sigma_x \hat{e}_x' + \sigma_y \hat{e}_y' + \sigma_z \hat{e}_z')$ where $\{\hat{e}_i\}$ and $\{\hat{e}_i'\}$ are arbitrary orthonormal bases.

When $\nvi \nparallel \nvii$ we can choose the basis for the first and second qubit in a symmetric way such that

\begin{equation}
\begin{split}
    % --- First System ---
    & \rm Q1\left\{
    \begin{aligned}
        &\hat{e}_z = \nvi \\
        &\hat{e}_x = \frac{\nvi \times \nvii}{\sqrt{1-(\nvi\cdot\nvii)^2}} \\
        &\hat{e}_y = \hat{e}_z \times \hat{e}_x = \frac{(\nvi\cdot\nvii)\nvi - \nvii}{\sqrt{1-(\nvi\cdot\nvii)^2}}
    \end{aligned}
    \right. 
    \\[10pt] % Adds vertical space between the systems
    % --- Second System ---
    & \rm Q2\left\{
    \begin{aligned}
        &\hat{e}_{z}' = \nvii \\
        &\hat{e}_{x}' = \hat{e}_x \\
        &\hat{e}_{y}' = \hat{e}_{z}' \times \hat{e}_{x}' = \frac{\nvi - (\nvi\cdot\nvii)\nvii}{\sqrt{1-(\nvi\cdot\nvii)^2}}
    \end{aligned}
    \right.
\end{split}
\end{equation}

With this, the single-qubit terms of the Hamiltonian are diagonal
\begin{equation}
    H = \frac{\muB}{2} |\gi\B| \sigma_z \otimes \mathbb{1}  + \frac{\muB}{2} |\gii\B| \mathbb{1}\otimes \sigma_z + \frac 1 4 \sum\limits_{i,j}J_{ij}\sigma_i\otimes\sigma_j\, ,
\end{equation}
and the coefficient $J_{ij}$ of the exchange interaction term $\sigma_i\otimes\sigma_j$ can be written as
\begin{eqnarray}
\begin{split}
&J_{xx} = J_0\frac{(\nvi\times\nvii)\R\SO(\nvi\times\nvii)}{1-(\nvi\cdot \nvii)^2}\\
&J_{yy} = J_0\frac{((\nvi\cdot\nvii)\nvi-\nvii)\R\SO(\nvi-(\nvi\cdot\nvii)\nvii)}{1-(\nvi\cdot \nvii)^2}\\
&J_{zz} = J_0\nvi\R\SO\nvii
\labelEq{Jij_diag}
\end{split}
\end{eqnarray}
Furthermore, the off-diagonal terms read
\begin{eqnarray}
\begin{split}
&J_{xy} = J_0\frac{(\nvi\times\nvii)\R\SO(\nvi-(\nvi\cdot\nvii)\nvii)}{1-(\nvi\cdot \nvii)^2}\\
&J_{yx} = J_0\frac{((\nvi\cdot\nvii)\nvi-\nvii)\R\SO(\nvi\times\nvii)}{1-(\nvi\cdot \nvii)^2}\\
&J_{xz} = J_0\frac{(\nvi\times\nvii)\R\SO\nvii}{\sqrt{1-(\nvi\cdot \nvii)^2}}\\
&J_{zx} = J_0\frac{\nvi\R\SO(\nvi\times\nvii)}{\sqrt{1-(\nvi\cdot \nvii)^2}}\\
&J_{yz} = J_0\frac{((\nvi\cdot\nvii)\nvi-\nvii)\R\SO\nvii}{\sqrt{1-(\nvi\cdot \nvii)^2}}\\
&J_{zy} = J_0\frac{\nvi\R\SO(\nvi-(\nvi\cdot\nvii)\nvii)}{\sqrt{1-(\nvi\cdot \nvii)^2}}
\labelEq{Jij_offdiag}
\end{split}
\end{eqnarray}

The expressions above can be used even in the case of single-axis anisotropies: simply replace $J_0\R\SO$ in \refEq{Jij_diag}-\ref{eq:Jij_offdiag} with the more generic exchange matrix. Note that we get the same expression for $J_{zz}$ as Ref.~\cite{geyer_anisotropic_2024} along with the other elements of the exchange matrix. We now define the longitudinal and transverse exchange components as:
\begin{eqnarray}
\begin{split}
&\Jpara = J_{zz}\\
&\Jperp = [J_{xx} + J_{yy} + i(J_{xy}-J_{yx})]/2.\\
\labelEq{Jpara_Jperp}
\end{split}
\end{eqnarray}

By replacing in \refEq{Jij_diag}-\ref{eq:Jij_offdiag} in \refEq{Jpara_Jperp}, we can express $\Jperp$ as
\begin{eqnarray}
    \Jperp = \frac{J_0}{2}\frac{1}{1-(\nvi\cdot \nvii)^2}[\nvi\times\nvii-i((\nvi\cdot\nvii)\nvi-\nvii)]\cdot
    \R\SO[\nvi\times\nvii+i(\nvi-(\nvi\cdot\nvii)\nvii)]
\end{eqnarray}
Expanding and simplifying the equation above yields
\begin{eqnarray}
\begin{split}
    \Jperp = \cos(2\theta\SO)\frac{J_0+\Jpara}2 + i \frac{J_0}2 \sin(2\theta\SO) \nso\cdot(\nvi+\nvii)\\
    + J_0\sin^2(\theta\SO)\left[\frac{(\nso\cdot(\nvi-\nvii))^2+(\nso\cdot(\nvi\times\nvii))^2}{1-(\nvi\cdot\nvii)^2}\right.\\
    +\left.i\frac{(\nso\cdot(\nvi-\nvii))(\nso\cdot(\nvi+\nvii))}{1+\nvi\cdot\nvii} \right].
\end{split}
\end{eqnarray}

Hence, in first order in spin-orbit interaction ($\theta_{\rm SO}\ll90^{\circ}$), we find \refEq{J_first_order_main}:
\begin{eqnarray}
\begin{split}
\Jpara &= J_0\nvi\cdot\nvii\\
|\Jperp|&= (J_0+J_\parallel)/2
\labelEq{J_first_order}
\end{split}
\end{eqnarray}
which is in good agreement with \refFig{simulations}.

% \begin{eqnarray}
% H_{J} = H_{\parallel} + H_\perp = \frac{1}{4}\Jpara \sigma_{1}^{z} \sigma_2^{z} + \frac{1}{2} (J_\perp \sigma_1^+ \sigma_2^- + \text{h.c.})
% \end{eqnarray}

\begin{table}[ht]
    \centering
    \renewcommand{\arraystretch}{1.3} 

    % Define vertical lines in the preamble: |c|c|c|
    \begin{tabular}{|c|c|c|} 
        \hline
        \textbf{Param.} & \multicolumn{2}{c|}{\textbf{Value}} \\ \hline 
        
        % Global Parameters
        $J_0$~(MHz) & \multicolumn{2}{c|}{26.4(6)} \\
        $\:\:\theta_{S0}$~($^{\circ}$) & \multicolumn{2}{c|}{11.5(14)} \\ 
        $\alpha_{n_{S0}}$~($^{\circ}$) & \multicolumn{2}{c|}{\num{240(5)}} \\
        $\beta_{n_{S0}}$~($^{\circ}$) & \multicolumn{2}{c|}{\num{34(3)}} \\
        \hline
        
        % Split Parameters (Q1 vs Q2)
         & \textcolor{Q1brown}{\textbf{Q1}} & \textcolor{Q2pink}{\textbf{Q2}} \\
        \hline
        
        \textcolor{red}{$g_{x'}$} & \num{0.0697(5)}   & \num{0.0100(9)} \\
        \textcolor{blue}{$g_{y'}$} & \num{0.2697(10)}  & \num{0.2456(8)} \\
        \textcolor{green!50!black}{$g_{z'}$} & \num{12.75(11)} & \num{11.53(6)} \\
        \hline
        
        $\zeta_g$~($^{\circ}$) & \num{33.9(14)}  & \num{-16.8(6)} \\ 
        $\theta_g$~($^{\circ}$) & \num{0.479(7)} & \num{0.819(3)} \\ 
        $\phi_g$~($^{\circ}$) & \num{12.5(13)}  & \num{64.5(6)} \\   
        \hline
    \end{tabular}
    \caption{Fit parameters. Global parameters are listed at the top, followed by the parameters for Q1 and Q2 in separate columns.}
\labelSuppTab{fit_param}
\end{table}

\section{\labelSec{met:fitting} Fitting parameters resulting from the transition fitting}
%change title if you want

We define the fitting function as an array of transitions
\begin{eqnarray}
    {\bf f}_\text{fit}({\bf p},\theta,\phi) = \{\Delta_{1,2},\,
    \Delta_{3,1},\,
    \Delta_{4,2},\,
    \Delta_{4,3},\,
    \Delta_{4,1}\}
    \label{eq:trans_fitt_func}
\end{eqnarray}
where $\bf{p}$ is an array of 16 fitting parameters including the g-tensors of the two qubits (2$\times$6 parameters), the scalar exchange $J_0$, the spin-orbit angle $\theta\SO$ and the unit spin-orbit vector $\nso(\alpha_{n_{SO}},\beta_{n_{SO}})$ (1+1+2 parameters). Moreover, $\Delta_{i,j} = (\varepsilon_i-\varepsilon_j)/h$ are the 5 measured transition calculated from the eigenvalues $\varepsilon_i$ of the Hamiltonian
Eq.~\eqref{eq:Hlab} for a given set of parameters ${\bf p}$ and magnetic field direction $\B(\theta_B,\phi_B) = B(\sin\theta_B \cos\phi_B,\sin\theta_B \sin\phi_B,\cos\theta_B)$.

The fitting function is then compared to the datasets presented in \refSuppFig{transitions} ordered as
\begin{eqnarray}
\begin{split}
    {\bf f}(\theta_B,\phi_B) = \{&\min(f_{Q1,\downarrow},f_{Q2,\downarrow}),\,
    \max(f_{Q1,\downarrow},f_{Q2,\downarrow}),\,\\
    &\max(f_{Q1,\uparrow},f_{Q2,\uparrow}),\,
    \min(f_{Q1,\uparrow},f_{Q2,\uparrow}),\,\\
    &f_{\uparrow\uparrow,\downarrow\downarrow}\}
    \label{eq:trans_fitt_func}
\end{split}
\end{eqnarray}
Apart from the 2D scans ($\leq225$ data points) from \refFig{spectroscopy} the data list was extended with data points measured in the mean plane ($25$ data points). Data points where the transition was not visible were taken into account with zero weight in the least-square fit.

The initial values for the g-tensors were obtained from fitting the extrapolated qubit-frequencies (derived from the qubit frequencies at zero barrier voltage and the measured frequency derivatives). The fit parameters of the g-tensors were constrained to the range of $(80\%,120\%)$ of the initial value. Relaxing the constraints did not change the obtained fitting values but did increase the time of the optimisation. The best fitting parameters are presented in \refSuppTab{fit_param}.

We performed least-square fitting with and without ($\theta\SO=0$) spin-orbit induced spin rotations and for the case of non-zero $\theta\SO$ we did not find optimal parameters that were robust to changing initial parameters. We therefore conclude that the spin-orbit induced spin rotations are too small to be reliably determined from this experiment. 

% \begin{figure*}[htp]
% \includegraphics[width=146.8mm]{Fig_device_cry.pdf}
% \caption[Device and experimental setup]{\textbf{Device and experimental setup}: 
%     False-coloured SEM-image of the used DQD+SHT, with different colours for each layer, as shown in the legend.
%     The white reference frame shows the orientation of the device in the lab frame, used to define the magnetic field angles and components.
%     Furthermore, the current-based readout setup is schematised. The current, converted into a voltage, is then thresholded and translated into state probability, as detailed in \refSuppFig{threshold}.
%     }
% \labelSuppFig{device}
% \end{figure*}

\section{\labelSec{met:QPT} Quantum process tomography of the baseband \MakeLowercase{i}SWAP and SPAM-budget analysis}

\subsection{Tomography protocol}

We perform standard linear-inversion QPT in an informationally-complete 16-state basis. For each two-qubit input state $|\psi_j\rangle$, $j = 1, \ldots, 16$, formed as a tensor product of single-qubit states drawn from $\{|0\rangle, |1\rangle, |+\rangle, |i\rangle\}$, we apply the iSWAP gate and then read out by projecting onto one of 16 measurement basis states $\{|\psi_k\rangle\}$ via the corresponding inverse single-qubit rotation followed by readout. The single-qubit gates used at the $J$-off point are $X_\pi$ (for $|1\rangle$), $Y_{\pi/2}$ (for $|+\rangle$) and $X_{-\pi/2}$ (for $|i\rangle$). The resulting $16\times 16$ probability matrix is $P_\mathrm{iSWAP}[j,k] = \mathrm{Tr}(|\psi_k\rangle\langle\psi_k|\, \mathrm{iSWAP}|\psi_j\rangle\langle\psi_j|)$, with each entry estimated from $N_\mathrm{shots}=1000$ binary single-shot outcomes thresholded as in \refSuppFig{threshold}. Subsequentially, we acquire the SPAM matrix $P_\mathrm{SPAM}$ from the same protocol with the identity $I$ (idling for $t_{J,\rm iSWAP}$) in place of the iSWAP pulse, satisfying $P_\mathrm{SPAM}[j,k] = \mathrm{Tr}(|\psi_k\rangle\langle\psi_k|\, I(t_{J,\rm iSWAP})|\psi_j\rangle\langle\psi_j|)$. The outcomes of the measurements are plotted in \refSuppFig{FullQPT}.

\subsection{Choi reconstruction and fidelity}

From $P_\mathrm{iSWAP}$ we reconstruct the Choi matrix by linear inversion. Defining the operator-overlap matrix $O[i,j] = \mathrm{Tr}(\rho_i \rho_j) = |\langle\psi_i|\psi_j\rangle|^2$ for the ideal prep states $\rho_j = |\psi_j\rangle\langle\psi_j|$, the action of the channel on each prep state decomposes as $\mathcal{E}(\rho_j) = \sum_k C[j,k]\, \rho_k$ with $C = P_\mathrm{iSWAP} \, O^{-1}$. The Choi matrix is then assembled in the convention $\chi = \sum_{a,b} |a\rangle\langle b| \otimes \mathcal{E}(|a\rangle\langle b|)$, with $|a\rangle\langle b|$ decomposed in the $\{\rho_k\}$ basis via $O^{-1}$. In this convention $\mathrm{Tr}(\chi) = d = 4$ for a trace-preserving channel and $\mathrm{Tr}(\chi^2) = d^2$ for a unitary. The reconstructed $\chi_\mathrm{raw}$ is in general not physical (its eigenvalues may be negative and its trace may deviate from $d$) and we project it onto the closest physical Choi matrix in Frobenius distance ($||A-B||_F = \sqrt{\Sigma_{i=1}^n \Sigma_{j=1}^m |a_{ij}-b_{ij}|^2}$, where $A,B$ are $n\times m$ matrices) using the closed-form algorithm of Ref.\cite{smolin_efficient_2012}, which returns the closest positive semidefinite matrix with fixed trace. The process fidelity to an ideal unitary $U$ is then $F_\mathrm{QPT} = \mathrm{Tr}(\chi_U^\dagger \chi_\mathrm{phys})/d^2$, where $\chi_U = \sum_{a,b} |a\rangle\langle b| \otimes U|a\rangle\langle b| U^\dagger$.

\subsection{SPAM and estimation of $F_\mathrm{iSWAP}$}
The standard linear SPAM correction $P_\mathrm{corr} = P_\mathrm{gate}\, P_\mathrm{SPAM}^{-1}\, P_\mathrm{SPAM,ideal}$, which models all SPAM error as a readout distortion to be undone by inversion, produces unbiased results only when $P_\mathrm{SPAM}$ is well-conditioned. Given the poor SPAM at the iSWAP magnetic field direction (the causes are explained in the next subsection),
we therefore report $F_\mathrm{QPT} = 60\%$ computed directly from $P_\mathrm{iSWAP}$ without SPAM correction, as a conservative figure-of-merit that includes SPAM contamination.

Approximating each SPAM operation as a depolarising channel $\mathcal{D}_p(\rho) = (1-p)\,\rho + \frac{p}{d}\,\mathbb{1}$ acting on a 2-qubit state (where $p$ is the depolarising error probability)\cite{nielsen_simple_2002}, the round-trip fidelity for any pure state is $\langle\psi|\,\mathcal{D}_p(|\psi\rangle\langle\psi|)\,|\psi\rangle = 1 - \frac{3}{4}p$ (using $d = 4$). The full SPAM round-trip composes a preparation depolarisation and a measurement depolarisation (with associated error probabilities $p_p$ and $p_m$, respectively):

$$
P_\mathrm{SPAM}[j,j] \approx \left(1 - \frac{3}{4}p_p\right)\left(1 - \frac{3}{4}p_m\right).
$$

Averaging the measured diagonal over the 16 basis states gives $\langle\mathrm{diag}(P_\mathrm{SPAM})\rangle = 0.69$. Under the symmetric ansatz $p_p = p_m = p$, this yields $p \approx 0.21$ per SPAM operation. The measured QPT fidelity is approximately the product of the gate fidelity and the SPAM round-trip:

$$
F_\mathrm{QPT} \approx F_\mathrm{iSWAP} \cdot \left(1 - \frac{3}{4}p_p\right)\left(1 - \frac{3}{4}p_m\right) = F_\mathrm{iSWAP} \cdot \langle\mathrm{diag}(P_\mathrm{SPAM})\rangle,
$$

and inverting:

$$
F_\mathrm{iSWAP} \approx \frac{F_\mathrm{QPT}}{\langle\mathrm{diag}(P_\mathrm{SPAM})\rangle} \approx \frac{0.60}{0.69} \approx 0.87.
$$

We emphasise that this is a back-of-envelope estimate that (i) assumes depolarising SPAM, whereas the actual errors contain coherent components (X-rotation under- or over-shoots, Z-phase accumulation during finite-duration pulses); (ii) assumes prep and measurement errors are equal; (iii) does not address the gauge freedom that a full gate-set tomography would expose. This estimate is meant to set the scale of the underlying gate fidelity rather than as a definitive figure of merit.

\subsection{Causes of poor SPAM at the iSWAP point}

The iSWAP magnetic-field orientation $(\phi_{B,\mathrm{iSWAP}}, \theta_{B,\mathrm{iSWAP}})$ is fixed by the simultaneous conditions $J_\parallel = 0$ and $\Edg = 0$, which from the fits in \refSec{iSWAP} and \refFig{simulations}, place the field outside the equatorial $x'y'$-planes of both $g$-tensors and close to the $\hat g_{x'}$ directions. 
Out of the equatorial plane, the qubits are no longer at the hyperfine-noise sweet spot characterised in \refSec{coherence} and in Ref.\cite{hendrickx_sweetspot_2024,yu_optimising_2026}. Furthermore, the at the $\hat g_{x'}$ directions, the $g$-tensors vary sharply with $(\phi_B, \theta_B)$, implying faster driving, but increased charge-noise coupling.
This results in a substantially shorter $T_2^*\sim1.5$~$\mu$s (compared to the $T_2^*=21.3$~$\mu$s measured at $\hat g_{y'}$), so phase errors accumulate over the SPAM circuit duration. These effects compound especially for the more demanding basis states that require concatenated rotations on both qubits, e.g.\ $|1+\rangle$ needs $X_\pi$ on Q1 plus $Y_{\pi/2}$ on Q2.
Developments in novel germanium-based heterostructures have the promise to solve these problems. Unstrained heterostructures\cite{costa_buried_2026} can result in less strain-dominated $g$-tensors with higher voltage tunability, that combined with isotopic purification\cite{zeng_highfidelity_2026, daoust_nuclear_2026}, can be used to engineer $g$-tensors that allow for $\B$ orientations with simultaneous high coherence, fast single-qubit operations via hopping\cite{wang_operating_2024}, and favourable combinations of $\Jpara$, $\Jperp$ and $\Edg$.

\begin{figure}[h!]
\includegraphics[width=140mm]{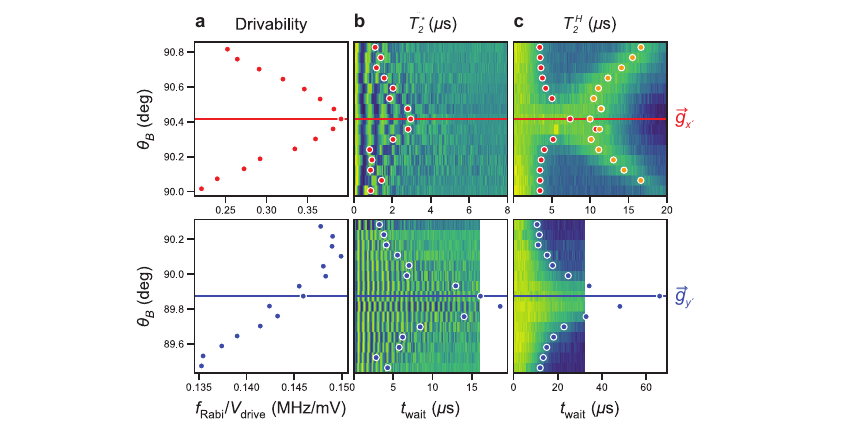}
\caption[Comparison of drivability and coherence times aligned with $\vec{g}_{x'}$ and $\vec{g}_{y'}$ at fixed qubit Larmor frequency - Q1]{\textbf{Comparison of drivability and coherence times aligned with $\vec{g}_{x'}$ and $\vec{g}_{y'}$ at fixed qubit Larmor frequency $f_{\rm Q1} \sim100$~MHz}: 
\textbf{a},~Drivability defined as the Rabi frequency $f_{\rm Rabi}$ normalised by the driving amplitude $V_{\rm drive}$, plotted as a function of the magnetic field angle $\theta_B$. The top (bottom) panel corresponds to a angular sweep around the $\vec{g}_{x'}$ ($\vec{g}_{y'}$) axis.
\textbf{b},~Ramsey coherence time $T_2^*$. The background colour map displays the raw measurement traces (Ramsey fringes), while the dots indicate the fitted decay times.
\textbf{c},~Hahn echo coherence time $T_2^H$, with raw traces in the background and fitted decay times as dots. The orange dots in the top panel indicate the spin population revival induced by the $^{73}$Ge nuclear-spin bath.
In all panels, the horizontal solid lines (red for $\vec{g}_{x'}$, blue for $\vec{g}_{y'}$) indicate the angle of the principal axes. Both drivability and coherence times exhibit a local maximum at these angles.
}
\labelSuppFig{T2_vs_theta}
\end{figure}

\begin{figure}[h!]
\includegraphics[width=140mm]{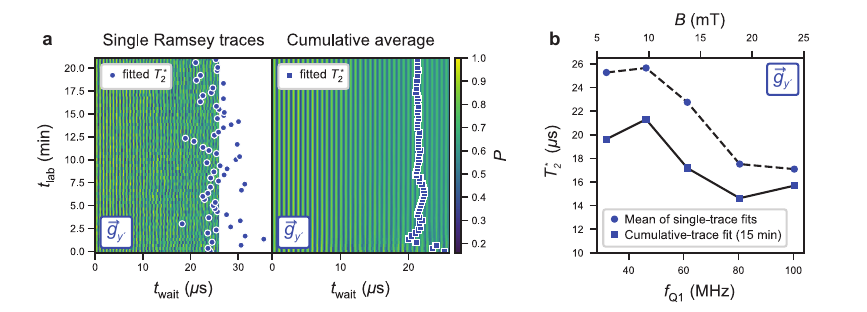}
\caption[Ramsey coherence time in the ergodic limit at $\vec{g}_{y'}$ - Q1]{\textbf{Ramsey coherence time of Q1 in the ergodic limit at $\vec{g}_{y'}$}: 
\textbf{a},~Free induction decay (Ramsey) traces as a function of wait time $t_{\rm wait}$, recorded continuously over 20~minutes of laboratory time $t_{\rm lab}$ at $f_{\rm Q1} = 45$~MHz. The left panel displays single-shot traces and their fitted decoherence times $T_2^*$ (blue dots). The right panel shows the cumulative average of the traces and the resulting fitted $T_2^*$ values (blue squares).
\textbf{b},~Ramsey decoherence time as a function of qubit Larmor frequency $f_{\rm Q1}$ and magnetic field magnitude $B$. Blue dots indicate the mean of the single-trace $T_2^*$ values (shown in \textbf{a}, left), while blue squares indicate the $T_2^*$ extracted from a 15-minute cumulative average decay.
}
\labelSuppFig{T2_ergodic}
\end{figure}

\begin{figure}[h!]
\includegraphics[width=140mm]{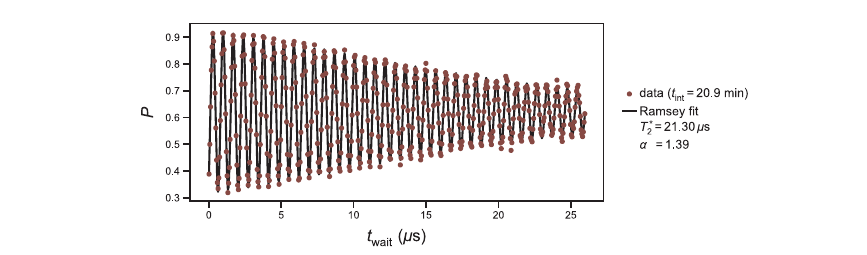}
\caption[Ramsey free induction decay in the ergodic limit at $\vec{g}_{y'}$]{\textbf{Ramsey free induction decay of Q1 in the ergodic limit at $\vec{g}_{y'}$}: The data, which results from averaging all the traces of \refSuppFig{T2_ergodic}a (total $t_{\rm int}=20.9$~min), is fitted with $P(t_{\rm wait})\propto \exp[-(t_{\rm wait}/T_{2}^*)^\alpha]$. The qubit Larmor frequency is $f_{\rm Q1} = 46.3$~MHz for a magnetic field magnitude $B = 9.7$~mT.
}
\labelSuppFig{T2_last_ergodic_fit_45MHz}
\end{figure}

\begin{figure}[htp]
\includegraphics[width=125.6mm]{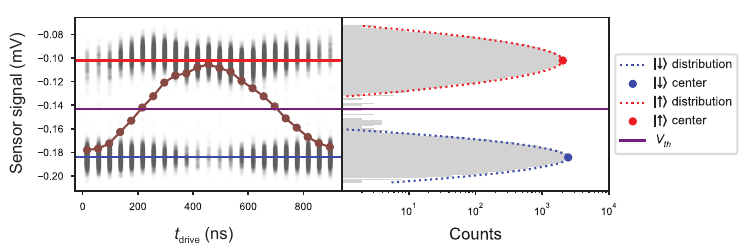}
\centering
\caption[Single-shot readout and signal thresholding]{\textbf{Single-shot readout and signal thresholding}:
    Left panel: The sensor signal plotted as a function of Q1 Rabi drive duration $t_{\text{drive}}$. The background cloud (grey points) consists of individual single-shot measurement outcomes. The brown dots represent the averaged measured signal at each time step, displaying a full period of Rabi oscillation. 
    Right panel: Histogram of the sensor signal counts. The data reveals a bimodal distribution corresponding to the two spin states, which are fitted by two Gaussians (dotted lines). The horizontal solid lines indicate the centres of the ground state $\kd$ (blue) and excited state $\ku$ (red) distributions. Their midpoint (purple line) marks the discrimination threshold voltage $V_{th}$ used to translate the readout signal from voltage to spin-state probability.}
\centering
\labelSuppFig{threshold}
\end{figure}

\begin{figure*}[htp]
\includegraphics[width=146.8mm]{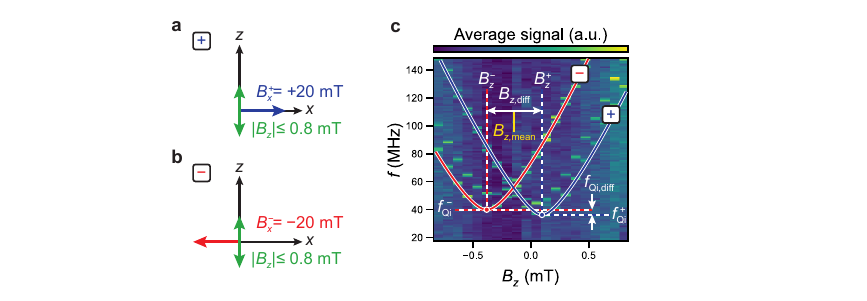}
\caption[Magnetic-field offset calibration]{\textbf{Magnetic-field offset calibration}: 
    \textbf{a},~Scheme of the Cartesian components of the magnetic field applied during the first spectroscopic measurement "+", with fixed positive $B_X^+$ (blue) and swept $B_z$ (green). 
    \textbf{b},~Scheme of the Cartesian components of the magnetic field applied during the second spectroscopic measurement "-", with fixed negative $B_X^-$ (red) and swept $B_z$ (green). 
    \textbf{c},~Mean of the signal obtained during the two spectroscopic measurements. The solid blue and red parabolic lines track the frequency of qubit Qi as a function of $B_z$ for measurement "$+$" and "$-$", respectively. The coordinates $(B_z^{\pm}, f^{\pm}_{\rm Qi})$ of the frequency minima (coloured dots) are used to extract $B_{z,\rm{offset}} = B_{z,\rm{mean}}$. In each spectroscopic measurement, two distinct frequency branches are visible, associated to two different qubits. However, this calibration protocol only requires one.
    }
\labelSuppFig{calibB_field}
\end{figure*}

\begin{figure*}[htp]
\includegraphics[width=120mm]{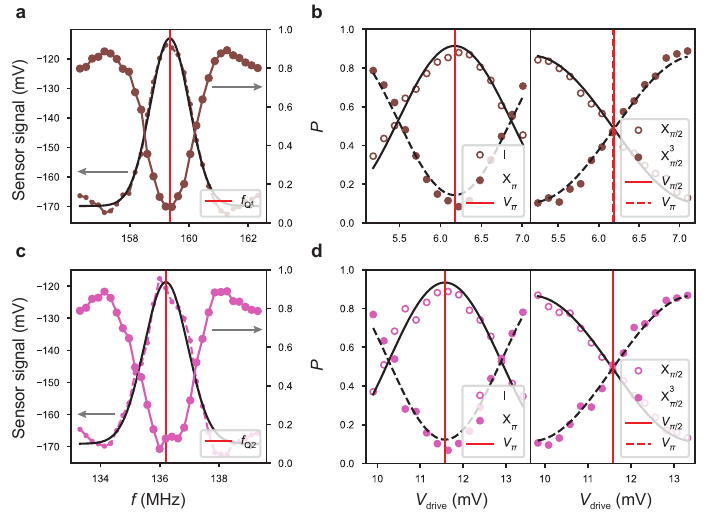}
\caption[Three calibration steps of AutoCalib routine]{\textbf{Three calibration steps of AutoCalib routine}: 
    \textbf{a}~and \textbf{c}, ~Qubit frequency calibration (1$^{\rm st}$ step) for Q1 and Q2. The qubit frequency $f_{\text{Q}i}$ is extracted by fitting a Gaussian on the sensor signal (large dots). Small dots indicate the $\kd$ probability, obtained by thresholding the sensor signal. 
    \textbf{b}~and \textbf{d}, X$_\pi$ (left) and X$_{\pi/2}$ (right) gate calibration (2$^{\rm nd}$ and 3$^{\rm rd}$ steps) for Q1 and Q2. The number of repetition for the error-amplifying sequence is $n=1$ in this case, but can be chosen to be larger for higher accuracy.
    }
\labelSuppFig{autocalib}
\end{figure*}

\begin{figure*}[htp]
\includegraphics[width=146.8mm]{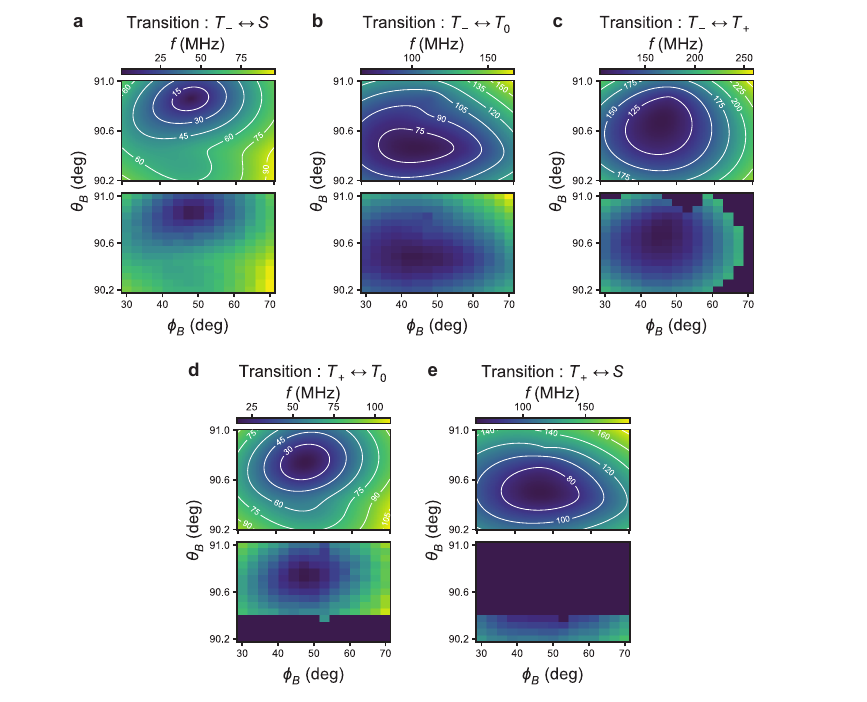}
\caption{\textbf{Comparison between simulated and measured transition frequencies as a function of magnetic field angle}: 
    Dependence on the magnetic field direction ($\phi_B, \theta_B$) of the simulated frequency (using the best fitting parameters shown in \refSuppTab{fit_param}, top panel) and measured frequency (via spectroscopy, bottom panel) of transition: 
    \textbf{a}, $T_-\leftrightarrow S$,
    \textbf{b}, $T_-\leftrightarrow T_0$,
    \textbf{c}, $T_-\leftrightarrow T_+$,
    \textbf{d}, $T_+\leftrightarrow T_0$,
    \textbf{e}, $T_+\leftrightarrow S$.
    }
\labelSuppFig{transitions}
\end{figure*}

\begin{figure*}[htp]
\includegraphics[width=180mm]{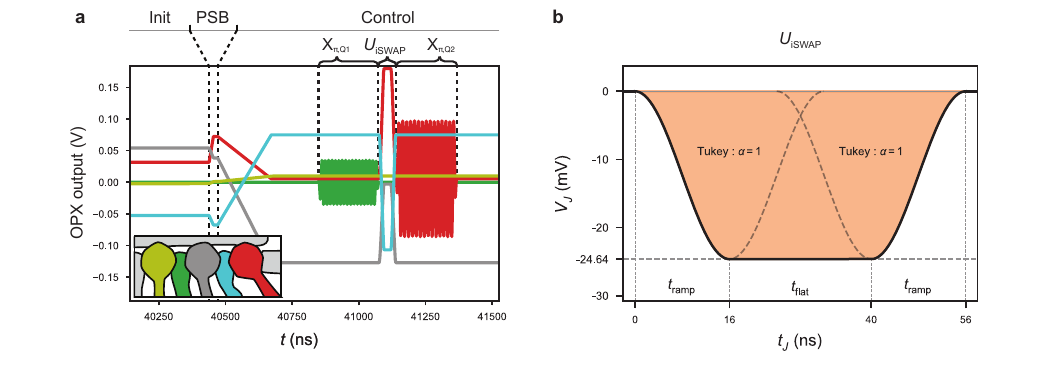}
\caption{\textbf{Baseband pulse sequence and iSWAP pulse shape}: 
    \textbf{a}, Time trace of the OPX output voltages on the device gates during a representative iSWAP experiment, comprising the initialisation (Init), Pauli-spin-blockade readout window (PSB), and control segment. The control segment shown here is an example sequence including a single-qubit $X_\pi$ pulse on Q1, the baseband exchange pulse $U_\mathrm{iSWAP}$, and an $X_\pi$ pulse on Q2, with the single-qubit microwave drives visible as the high-frequency bursts and the exchange pulse as the sharp baseband excursion on the barrier and plunger gates. Coloured traces correspond to the individual gates, colour-keyed to the device schematic in the inset (plungers and barrier). 
    \textbf{b}, Detail of the calibrated exchange pulse $U_\mathrm{iSWAP}$ applied to the virtual exchange voltage $V_J$. The pulse is a Hann window (Tukey window with shape parameter $\alpha = 1$) composed of $t_\mathrm{ramp} = 16$~ns cosine rise and fall, a $t_\mathrm{flat} = 24$~ns flat top at the calibrated amplitude $V_{J,\mathrm{iSWAP}} = -24.64$~mV, for a total duration $t_\mathrm{iSWAP} = 56$ ns. Dashed grey curves indicate the rising and falling cosine tapers.
    }
\labelSuppFig{iSWAP_pulse}
\end{figure*}

\begin{figure*}[htp]
\includegraphics[width=180mm]{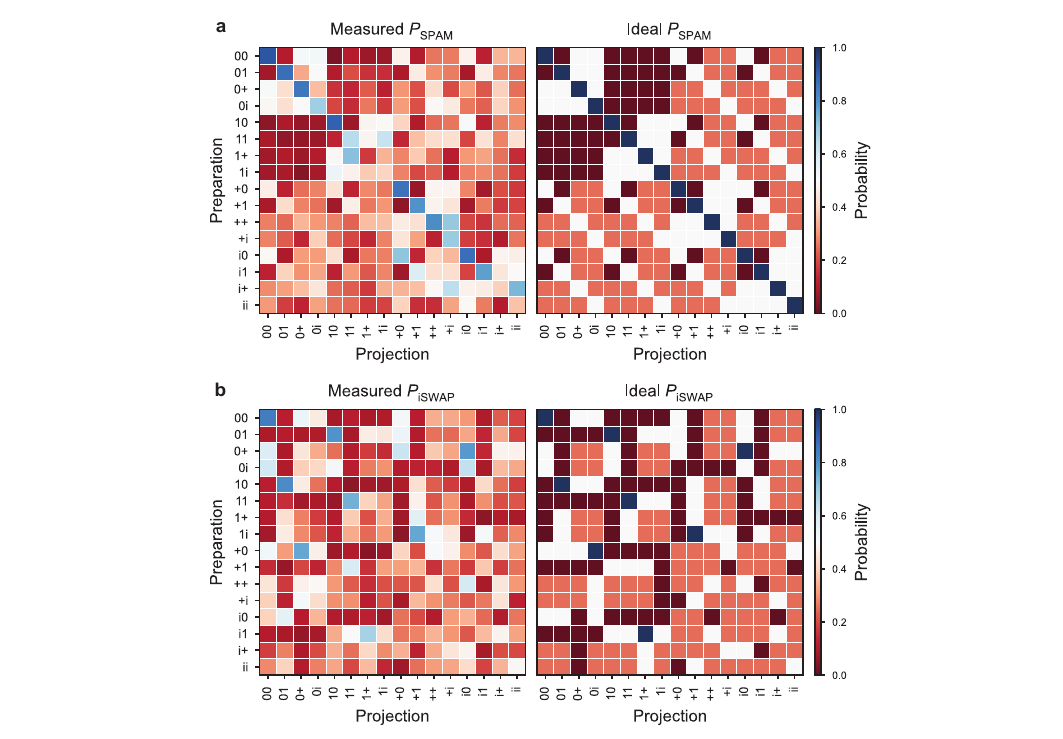}
\caption{\textbf{Full 16-state quantum process tomography and SPAM characterisation of the baseband iSWAP gate}: 
    \textbf{a}~ Measured (left) and ideal (right) SPAM probability matrices $P_\mathrm{SPAM}$, acquired by preparing and projecting each of the 16 basis states $\{\ket{\psi_i}\}_{i=1\ldots16}$ formed by all tensor products of $\{\ket{0}\coloneqq \kd, \ket{1}\coloneqq \ku, \ket{+}, \ket{i}\}$, with no gate applied (idling for $t_{J,\rm iSWAP} = 56$~ns). Matrix entries are $P_\mathrm{SPAM}[i,j] = |\bra{\psi_j}I\ket{\psi_i}|^2$. The ideal $P_\mathrm{SPAM}$ is not the identity: because the 16 states are non-orthogonal, the ideal entry is the geometric overlap $|\langle\psi_k|\psi_j\rangle|^2$, which is 1 on the diagonal, 0 between orthogonal pairs (e.g. $|0\rangle$ and $|1\rangle$), and $1/2$ or $1/4$ for superposition-state pairs (e.g. $|\langle 0|{+}\rangle|^2 = 1/2$), producing the structured block pattern visible in the right panel. Comparing the measured and ideal matrices reveals the reduction in diagonal entries for states requiring concatenated rotations on both qubits ($|1{+}\rangle$, $|1i\rangle$, $|i{+}\rangle$, $|ii\rangle$), quantified by the diagonal bar chart in main-text \refFig{iSWAP}b. The dashed box highlights the $4\times4$ computational-basis sub-block $\{|00\rangle,|01\rangle,|10\rangle,|11\rangle\}$, shown as the 3D bar plot in main-text \refFig{iSWAP}c.
    \textbf{b}~ Measured (left) and ideal (right) iSWAP probability matrices $P_\mathrm{iSWAP}$, obtained from the same protocol with the calibrated baseband iSWAP pulse ($t_\mathrm{iSWAP} = 56$ ns, $V_{J,\mathrm{iSWAP}} = -24.64$ mV) in place of the identity. The ideal matrix encodes the iSWAP probability amplitudes $P_\mathrm{iSWAP}[j,k] = |\langle\psi_k|\mathrm{iSWAP}|\psi_j\rangle|^2$ over the full non-orthogonal basis; its non-trivial off-diagonal structure arises from the superposition-state overlaps of the basis combined with the state-swapping action of the gate. The dominant features of the ideal pattern are recovered in the measurement, with reduced contrast due to SPAM errors. The process fidelity $F_\mathrm{QPT} = 60\%$ and the estimate iSWAP fidelity $F_\mathrm{iSWAP} = 87\%$ are extracted from the measured $P_\mathrm{iSWAP}$ following the reconstruction procedure described in \refSec{met:QPT}.
    }
\labelSuppFig{FullQPT}
\end{figure*}

%TC:endignore
\end{document}